# Evolution of Linear Viscoelasticity across the Critical Gelation Transition

Yogesh M. Joshi[1, 2, 3, *]

[1] Department of Chemical Engineering, Indian Institute of Technology Kanpur, Kanpur, Uttar Pradesh 208016, India.

[2] Materials Science Programme, Indian Institute of Technology Kanpur, Kanpur, Uttar Pradesh 208016, India.

[3] Centre for Nanosciences, Indian Institute of Technology Kanpur, Kanpur, Uttar Pradesh 208016, India.

* Email: joshi@iitk.ac.in

## Abstract

In this work, we develop a rigorous theoretical framework for the evolution of linear viscoelastic properties across the sol-gel transition. More specifically, we derive general admissible expressions for the relaxation modulus and dynamic moduli as the critical gel state is approached from the pre-gel or the post-gel side. These expressions possess a generalized multi-mode series representation and recover the critical gel power law spectrum in the limit of vanishing distance from the gel point. We validate these expressions against the experimental data for various polymeric and colloidal systems. A central finding of the present work is the requirement of continuity of the dynamic moduli and their derivatives at the critical gel point, which imposes a profound physical constraint, necessitating the relaxation dynamics on both sides of the transition to be symmetric. This, in turn, leads to the hyper-scaling relation, which is a theoretical requirement rather than an empirical proposal. We further show that the critical relaxation exponent ($n$) always remains above the relaxation scaling exponent ($\kappa$), establishing a previously unrecognized lower bound on $n$. We also analytically estimate, for the first time, the parameter $C$ that characterizes the relative evolution of the storage modulus with respect to the loss modulus as the critical state is approached. These results reveal that the symmetry, scaling, and hyperscaling properties of the sol-gel transition are all consequences of a single unifying physical requirement originating from the continuity of the linear viscoelastic properties at the critical gel point.

**Introduction**

Gels encompass a wide range of systems, from synthetic polymers to colloidal systems, and represent a unique state of soft matter characterized by a space-spanning percolated network.[1] Polymeric gels include covalently crosslinked or physically associated networks, while the colloidal gels are formed through particle aggregation, depletion, or electrostatic interactions.[2-3] Notwithstanding differences in the nature of network that constitute the gel, both types of systems show remarkable similarity in their rheological behavior.[4] The process by which a low-viscosity liquid (sol) transforms into a soft elastic solid (gel) is known as the sol-gel transition, which, depending on the system, could occur either spontaneously, by changing temperature, or by the addition of chemicals that control the extent of crosslinking. Understanding the rheological behavior of the sol-gel transition is not only of fundamental importance but also of immense practical relevance. Particularly, the ability to precisely track and control this transition is vital in diverse fields of applications, including the formulation of biomaterials,[5-7] coatings,[8] adhesives,[9] additive manufacturing or 3D printing,[10] food processing,[11] biomedical devices, energy materials,[12-13] and functional polymers,[14] etc. These applications rely on the ability to control the flow and solidification behavior, wherein rheology serves as the key diagnostic to connect microstructure with macroscopic performance, providing a window into the evolving network connectivity, relaxation spectrum, and timescales that define the sol–gel transition. Among the many milestones in this field, the seminal contribution by Winter and Chambon,[15] nearly four decades ago, introduced the concept of the critical gel state, a unique state between the sol and the gel, which established a mathematically precise rheological fingerprint characterized by self-similar relaxation and power-law viscoelasticity. However, while the critical gel itself has been extensively explored in the literature, an understanding of the evolution of viscoelastic properties as a material passes through the critical state is necessary to uncover the intrinsic, frequency-invariant relationships governing the network formation. In this work, we present a comprehensive analysis of this evolution and reveal previous unrecognized intrinsic insights that emerge from analysing the relative change in various linear viscoelastic properties.

To understand how linear viscoelastic properties evolve as the material transforms from the sol to the gel state, it is essential to understand the rheological characteristics of the critical gel state. The relaxation modulus at the critical gel state follows a power law dependence on time given by:[1]

$$G(t) = St^{-n} \qquad (1)$$

where $S$ is a quasi-property called gel strength, while $n$ is the critical relaxation exponent bounded by 0 and 1. The gel strength $S$ has units of $\text{Pa\,s}^n$, which for $n = 0$ reduces to a modulus, while for $n = 1$, it corresponds to viscosity. The Fourier transform of the relaxation modulus directly leads to the complex modulus, $G^*(\omega) = S\Gamma(1-n)(i\omega)^n$, which can be deduced to the storage and loss moduli expressed as:[1]

$$G' = S\Gamma(1-n)\cos(n\pi/2)\,\omega^n, \quad (2)$$
$$G'' = S\Gamma(1-n)\sin(n\pi/2)\,\omega^n,$$

where $\Gamma(1-n)$ is the Euler gamma function of $(1-n)$. On either side of the critical gel state lies the sol (pre-gel) and the post-gel state. The progress of the system as it transforms from the pre-gel to the post-gel state is quantified by the degree of crosslinking given by parameter $p$, which is defined as the fraction of cross-linkable units that participate in the crosslinks. The degree of crosslinking, therefore, is bounded by 0 and 1. The degree of crosslinking at the critical gel state is defined as $p_c$, and the 'distance' from the same is expressed as $\varepsilon = |p - p_c|$.

In the pre-gel state ($p < p_c$), the material is a viscous liquid notwithstanding $G''$ dominates over $G'$ or not. As the cluster or network formation advances, $p$ increases and eventually approaches $p_c$, the point at which the network percolates and spans the entire available space. This growth causes divergence of the longest relaxation time as well as the bulk viscosity such that both properties assume value of infinity at the critical gel state.[4] At the unique critical gel state, the system possesses self-similar hierarchical structure whose fractal dimension is given by:[16]

$$f_d = \frac{5(2n-3)}{2(n-3)} \quad (3)$$

The corresponding relaxation modulus and dynamic moduli are given by Eqs. (1) and (2). This condition implies that the loss tangent or $\tan\delta = G''/G'$ becomes independent of frequency. As the network build-up progresses further into the post-gel state ($p > p_c$), the percolated network consolidates, and the material develops a finite equilibrium modulus ($G_e$). The presence of non-zero $G_e$ means that the relaxation modulus and the storage modulus in the post-gel state no longer vanish respectively at infinite times and at zero frequency; instead, they approach $G_e$.

The rheological changes in the vicinity of the critical point adhere to the scaling and hyper-scaling laws, which relate the macroscopic phenomena to the fractal structure of the network. This theory predicts a power-law relationship for all the rheological properties that are sensitive to the connectivity of a material. As the pre-gel approaches the critical gel state, the steady-state viscosity ($\eta_0$) diverges, following a power law dependence given by:[1, 17-19]

$$\eta_0 \sim \varepsilon^{-s}, \quad (4)$$

and the corresponding longest relaxation time ($\tau_{max,S}$) scales as:[20]

$$\tau_{max,S} = \tau_S \varepsilon^{-1/\kappa_S}, \quad (5)$$

where $\tau_S$ is a constant of proportionality. Interestingly, there exists a hyper-scaling relationship between $\kappa_S$, $s$ and $n$ given by:[4, 21]

$$\kappa_S = \frac{(1-n)}{s}. \tag{6}$$

The relaxation modulus in the pre-gel state has been proposed to have a form:[20, 22]

$$G(t,\varepsilon) = St^{-n} \exp(-B_s \varepsilon t^{\kappa_S}), \tag{7}$$

which in the limit of the critical gel state $\varepsilon \to 0$ leads to the expression of the critical gel state.

On the other hand, once the material crosses into the post-gel state, the zero-frequency, or equilibrium modulus ($G_e$), begins to grow from zero, also following a power law dependence:[17, 19]

$$G_e(\varepsilon) = G_0 \varepsilon^z, \tag{8}$$

where $G_0$ is a material-specific constant. Once the space-spanning percolated network forms, the fraction of gel-forming species gradually decreases as the network consolidates. Consequently, the maximum relaxation time of the free or relaxable species ($\tau_{max,G}$) goes on decreasing with increase in $\varepsilon$ in the post-gel state, and is given by:[4, 21]

$$\tau_{max,G} = \tau_G \varepsilon^{-1/\kappa_G}, \tag{9}$$

where $\tau_G$ is the proportionality constant. Similar to the pre-gel state, there also exists a hyper-scaling relationship between $\kappa_G$, $z$ and $n$ given by:[4, 20]

$$\kappa_G = \frac{n}{z}. \tag{10}$$

Furthermore, for the post-gel state, Suman et al. recently proposed that the relaxation modulus is given by:[22]

$$G(t,\varepsilon) = St^{-n} \left( \sum_{m=0}^{\lfloor n/\kappa_G \rfloor} \frac{(B_g \varepsilon)^m t^{m\kappa_G}}{m!} \right) + G_e, \tag{11}$$

where $\lfloor n/\kappa_G \rfloor$ represents largest integer that is smaller than $n/\kappa_G$. A critical requirement for the consistency of various relationships are the hyper-scaling relations, which connects various power law coefficients with the critical exponent $n$, ensuring that the entire rheological spectrum is consistent at the critical point.

An intriguing aspect that emerges from experimental reports on various systems is that the relaxation dynamics evolve in an apparent symmetric manner on the two sides of the critical gel state, leading to: $\kappa_G = \kappa = \kappa_S$, which results in a hyperscaling relation:[1, 4, 22-23]

$$n = \frac{z}{z+s}. \tag{12}$$

Such symmetry suggests that the critical gel state may act as a dynamical fixed point, as the approach toward and beyond this state is governed by identical scaling exponents.

However, some systems do not report such symmetric evolution, implying $\kappa_G \neq \kappa_S$. Under what conditions such equality is observed, and under which it is not, remains an open question.

This work examines how the linear viscoelastic properties in the pre-gel state evolve as network formation progresses towards the critical gel state and beyond into the post-gel state. In this regard, Scanlan and Winter[20] experimentally observed that for a chemically crosslinking system approaching the critical gel state from the pre-gel state, the rate of change of storage and loss moduli with respect to $p$ shows a power-law dependence on $\omega$ given by:[18, 22]

$$\left(\frac{\partial \ln G^*}{\partial p}\right)_{p=p_c} \sim \left(\frac{\partial \ln G'}{\partial p}\right)_{p=p_c} = C_{S \to G} \left(\frac{\partial \ln G''}{\partial p}\right)_{p=p_c} \sim \omega^{-\kappa_S} \qquad (13)$$

The above relationship has been experimentally validated by several other groups. As noted by Winter and coworkers,[1, 21, 24-25] within experimental uncertainty $\kappa_S$ takes a universal value, around 0.2 for several dozen experiments on various polymers with different stoichiometry, chain length, and concentration. They observed that the value of $\kappa_S$ is insensitive to molecular details. Remarkably, Joshi and coworkers[26-29] reported the value of $\kappa_S$ for several colloidal systems undergoing sol-gel transition and report it to be around 0.2. For most of the systems studied in the literature, $\kappa_S$ has been observed to vary between 0.1 and 0.3. Winter and coworkers also reported the value of the parameter $C_{S \to G}$ to be around 2, which again was corroborated by Joshi and coworkers[26-29] for several colloidal systems. Recently, Morlet-Decarnin et al.[30] prepared various cellulose nanocrystals (CNCs) gels systems by varying CNC concentration as well as salt concentration and reported $\kappa_S$ to be in between 0.1 and 0.3 while $C_{S \to G}$ to be close to 2 for the same. While most of the work in the literature focuses on the sol-gel transition, Joshi and coworkers studied the gel-sol transition for various colloidal and polymeric systems, wherein the critical gel state was approached from the post-gel side. Interestingly, they also observed an identical relationship given by:[27, 31]

$$\left(\frac{\partial \ln G^*}{\partial p}\right)_{p=p_c} \sim \left(\frac{\partial \ln G'}{\partial p}\right)_{p=p_c} = C_{G \to S} \left(\frac{\partial \ln G''}{\partial p}\right)_{p=p_c} \sim \omega^{-\kappa_G}, \qquad (14)$$

wherein they reported the corresponding values of $\kappa_G$ to be around 0.2 and that of $C_{G \to S}$ to be around 2.

It is important to note that experimentally, the sol–gel transition is realized under different control modes. In many gel-forming systems, the transition occurs as a function of time, in which a liquid-like sol spontaneously evolves into a post-gel state, by passing the critical gel state due to chemical crosslinking reactions, aggregation, or structural arrest. In several systems the transition is driven by temperature, for example in thermos-reversible polymeric or physical gels, the sol–gel transition can often be obtained repeatedly by heating and cooling, provided the material does not undergo chemical degradation or aging.[4, 27, 31-33] For time-driven[25, 28, 30, 34-36] and thermally driven gelation, when the distance from the critical state is expressed respectively as $\varepsilon = |p - p_c| \sim |t - t_c|/t_c$ and $\varepsilon = |p - p_c| \sim |T - T_c|/T_c$, where $t_c$ and $T_c$ denote the corresponding values at the critical gel state, the system can be parametrized in terms of a single reduced distance variable. Interestingly, it is under these conditions, when the evolution toward the critical gel state is controlled by such a reduced distance, that scaling and hyperscaling relations have been observed to be obeyed. In other classes of materials, a completely different variable, such as the concentration of a reactive species, a crosslinker, or a dispersed phase, may serve as the primary control parameter that drives the sol-gel transition. In such cases, $\varepsilon$ is naturally defined in terms of the deviation from a critical concentration. Importantly, while the *physical origin* of the control parameter may differ across various systems, the rheological evolution near the critical gel state is often found to obey similar scaling forms, suggesting universality in the underlying viscoelastic response.

The sol–gel transition is traditionally investigated experimentally using small-amplitude oscillatory shear measurements performed on rotational rheometers leading to storage and loss moduli over an accessible frequency range. More recently, microrheological and particle-tracking techniques have been employed as complementary experimental methods that probe mean-squared displacement over extended time windows, enabling the measurement of local viscoelastic dynamics.[37-39] These particle-tracking methods not only lead to precise estimation of the critical gel state but also result in scaling and hyperscaling relations. These studies have largely confirmed critical scaling predictions and assessed issues such as the symmetry of relaxation dynamics and the universality of scaling exponents, thereby motivating further theoretical development.[38-43] Not just linear measurements, but also nonlinear viscoelastic experiments on systems undergoing a sol-gel transition lead to a better understanding of the behavior of interparticle/molecular bonds in the percolated network.[44-49] The gelation transition has also been studied theoretically through simulations and nonequilibrium statistical mechanics for polymeric as well as attractive colloidal systems, which give microscopic insights into this intriguing phenomenon.[50-56] The relaxation dynamics as a material undergoes the gelation transition have also been investigated by fractional viscoelasticity.[34, 36, 46, 57-58] Recently, machine learning based techniques have also been employed to study the sol-gel transition.[59]

With the above background, the primary motivation of the present work is to develop a systematic and unified understanding of how linear viscoelastic properties evolve as a material traverses from a pre-gel state to a post-gel state through the critical gel point, and vice versa. More specifically, we address the following issues:

1. An important issue that we would like to explore is the functional form of the relaxation modulus $G(t, \varepsilon)$. Although expressions such as Eqs. (7) and (11) have been proposed to describe the pre-gel and post-gel regimes, it remains unclear whether these forms are unique. If alternative representations are admissible, it becomes important to identify the constraints on the same.

2. Furthermore, the validity of hyper-scaling relations across the full spectrum of critical exponents warrants careful examination. While these relations are well supported for intermediate values of $n$, experimental studies probing the limit $n \to 0$ remain scarce. Assessing whether existing scaling frameworks remain valid over the entire spectrum of $0 < n < 1$, also constitutes another key objective of the present study.

3. As discussed above, some systems experimentally show that relaxation dynamics diverge from either side of the critical gel state in a symmetric manner, leading to: $\kappa_S = \kappa_G$, which then leads to hyper-scaling relation given by Eq. (12).[1, 4, 20-21, 24, 27, 31] In addition, theoretical treatments in the literature posit a symmetric divergence of the relaxation time across the critical gel transition.[23, 60-62] In addition, the exponent $\kappa$, on the sol or the gel side, has typically been reported to have a value close to 0.2 (between 0.1 and 0.3). The exponent $\kappa$ controls how this scale-free critical gel state characterized by exponent $n$ is approached from either side of the transition. An intriguing and unresolved issue is what dynamical constraints must a system satisfy to have $\kappa_G$ and $\kappa_S$ to be identical? Furthermore, it is important to explore what are the implications of this symmetry for the interpretation of linear viscoelastic data near the critical state.

4. As discussed above, the parameters $C_{S \to G}$ and $C_{G \to S}$ assume values close to 2 across a range of materials. These parameters may be interpreted as the logarithmic derivative $\left(\frac{\partial \ln G'(\omega,p)}{\partial \ln G''(\omega,p)}\right)_{p \to p_c}$, evaluated as the critical gel state is approached from either side. This implies that near the critical state, the logarithmic rate of change of the storage modulus is generically twice that of the loss modulus. This raises a fundamental question: does this derivative indeed assume a universal value of 2? If not, what are the conditions associated with the deviation.

**II. Model Development:**

In order to assess the evolution of linear viscoelasticity as a material transforms from a sol (pre-gel) state to a post-gel state, it is important to derive a generic form of relaxation modulus in both states as a function of time, as well as distance from the critical gel state $(G(t, \varepsilon))$. As mentioned in the previous section, there do exist a few proposals for the relaxation modulus for both states in the literature. However, in this section, we shall explore the possibility of finding a general expression and obtain various linear viscoelastic properties for the same. For the pre-gel and post-gel states, any acceptable form of the relaxation modulus must satisfy certain conditions mentioned in Table 1. We first discuss the pre-gel state, followed by the post-gel state.

**Table 1.** Conditions for any acceptable form of relaxation modulus in the pre- and post-gel states

| 1 | $G(t, \varepsilon) = 0 (t < 0), G(t, \varepsilon) \geq 0 (t > 0)$ |
|---|---|
| 2 | As $\varepsilon \to 0^+$ $$G(t, \varepsilon) \to S\, t^{-n}, 0 < n < 1$$ |
| 3 | For constant $\varepsilon$ $$\left(\frac{\partial G(t, \varepsilon)}{\partial t}\right)_\varepsilon \leq 0$$ |
| 4 | For all $\omega > 0$ the Fourier transform must satisfy: $$G'(\omega, \varepsilon) \geq 0, G''(\omega, \varepsilon) \geq 0$$ |
| 5 | For the pre-gel states $$G(t, \varepsilon) \to 0, \text{ for } t \to \infty$$ |
| 6 | For the post-gel states $$G(t, \varepsilon) \to G_e, \text{ for } t \to \infty$$ |

**II.1 Pre-gel state**

Considering these constraints, we propose the following generic relationship for the relaxation modulus:

$$G(t, \varepsilon) = St^{-n} \Phi\left(\frac{t}{\tau_{max,S}}\right), \quad (15)$$

where $\tau_{max,S}$, the characteristic relaxation timescale associated with the pre-gel state, is given by Eq. (5): $\tau_{max,S} = \tau_S\, \varepsilon^{-1/\kappa_S}$. A condition for the cut-off function $\Phi(x)$ based on the constraints mentioned in Table 1 is given by:

$$\Phi(x) = \begin{cases} 1, & x \to 0 \\ 0, & x \to \infty \end{cases}, \quad (16)$$

where the scaling variable $x$ is defined as $x = t/\tau_{max,S} = \varepsilon^{1/\kappa_S}\, t/\tau_S$. Furthermore, an additional experimental observation regarding the dependence of logarithmic derivative of dynamic moduli on frequency in a limit of $\varepsilon \to 0$ given by Eq. (13) can be transformed to obtain an expression for relaxation modulus, which has also been independently validated through experiments, and is given by:[20]

$$\left(\frac{\partial \ln G(t,\varepsilon)}{\partial \varepsilon}\right)_{\varepsilon \to 0} = -B_s t^{\kappa_S}, \tag{17}$$

Taking derivative of logarithm of $G(t,\varepsilon)$ Eq. (15) and further manipulation leads to:

$$\frac{\partial \ln G(t,\varepsilon)}{\partial \varepsilon} = \frac{x}{\kappa_S \varepsilon} \frac{d\ln \Phi(x)}{dx} \tag{18}$$

The limit of $x \to 0$ is the same as $\varepsilon \to 0$, and in order to recover Eq. (17) in that limit $\Phi(x)$ must follow

$$\lim_{x \to 0} \left\{\frac{d\ln \Phi(x)}{dx}\right\} = -\kappa_S B_s \tau_S^{\kappa_S} x^{\kappa_S - 1}. \tag{19}$$

In general, for $d\ln \Phi(x)/dx = -f(x)$, with the limit of $x \to 0$ given by Eq. (19) and imposing $\Phi(x) = 0$ for $x \to \infty$, we get:

$$\begin{aligned}
\Phi(x) &= \exp\left[-\int_0^x f(u) du\right], \\
u &\to 0, f(u) = \kappa_S B_s \tau_S^{\kappa_S} u^{\kappa_S - 1} \\
x &\to \infty, \int_0^x f(u) du = \infty
\end{aligned} \tag{20}$$

Therefore, any function $f(u)$ that satisfies the expression and constraints mentioned in Eq. (20) is an acceptable form of the cut-off function $\Phi(x)$ to be incorporated in the expression of relaxation modulus.

If we assume that the asymptotic form given by: $f(u) = \kappa_S B_s \tau_S^{\kappa_S} u^{\kappa_S - 1}$ holds over the entire range of $u$, from $0$ to $\infty$, we get $\Phi(x) = \exp(-B_s \tau_S^{\kappa_S} x^{\kappa_S})$ that obeys all the constraints mentioned in Eq. 20, and hence it constitutes an admissible form. The corresponding expression for relaxation modulus takes a form:

$$G(t,\varepsilon) = St^{-n} \exp(-B_s \varepsilon t^{\kappa_S}), \tag{7}$$

The expression given by Eq. (7), originally proposed by Scanlan and Winter,[20] satisfies all the fundamental constraints outlined in Table 1 while capturing the experimentally validated asymptotic behavior near the critical gel state. However, while Eq. (7) represents a minimal model and is widely adopted, but it is not unique. As evident from the general framework outlined in Eq. (20), any function $f(u)$ that satisfies the mentioned asymptotic behavior as $u \to 0$ and ensures divergence of the integral in a limit of $u \to \infty$ would result in an acceptable form for the cutoff function $\Phi(x)$. There could be several functional forms that satisfy the prescribed constraints; however, a general form can be expressed in terms of a multi-mode exponential representation given by:

$$G(t,\varepsilon) = St^{-n} \sum_{l=1}^{N} w_{S,l} \exp(-B_{S,l} \varepsilon t^{\kappa_S}) = St^{-n} \sum_{l=1}^{N} w_{S,l} \left[\sum_{k=0}^{\infty} \frac{(-1)^k B_{S,l}^k \varepsilon^k t^{k\kappa_S}}{k!}\right], \tag{21}$$

where $N$ represents the number of modes, $w_{S,l} > 0$ are the weight factors associated with each mode that satisfy $\sum_{l=1}^{N} w_{S,l} = 1$, and $B_{S,l} > 0$ are mode-specific coefficients. For $N = 1$, Eq. (21) reduces to the minimal form given by Eq. (7). The second expression in Eq. (21) is the complete Taylor series without truncation of the exponential term, which is a more convenient form for performing various transformations as discussed below. The equality between the first and second expressions is exact if the series is not truncated. In practice, only a finite number of terms in the series are retained (upper limit on $k$ is finite). In that case, the expansion is useful only when the argument of the exponential is less than unity: $B_{S,l}\varepsilon t^{\kappa_S} < 1$ for all $l$, which leads to the limiting condition: $t < (B_{S,l}\varepsilon)^{-1/\kappa_S}$. The multi-mode representation (Eq. (21)) constitutes a general framework for the pre-gel state relaxation modulus that adheres to all the prescribed constraints. This expression is also similar in structure to the well-established Prony series in linear viscoelasticity, which has long been recognized as a complete basis set capable of approximating any arbitrary relaxation function to the desired accuracy for a suitable choice of the number of modes. Eq. (21) too, for an appropriate choice of $N$ and the parameters $\{w_{S,l}, B_{S,l}\}$ can capture any pre-gel relaxation behavior that adheres to the prescribed constraints in series solution subject to $t < (B_{S,l}\varepsilon)^{-1/\kappa_S}$.

An essential feature of the sol-gel transition is the observation of time-cure superposition[1, 18, 30, 34, 38-39, 58, 63-66] that necessitates a change in degree of crosslinking $\varepsilon$ to affect only the mean value of relaxation time and not the shape of the relaxation time spectrum.[22, 66] For the pre-gel state, this means that the relaxation modulus should be expressible in the form $G(t, \varepsilon) = St^{-n}\Phi(t/\tau_{max,S})$, wherein dependence of $\varepsilon$ comes solely from the cutoff function $\Phi$ through $\tau_{max,S}$ without affecting the shape determined by $\{w_{S,l}, B_{S,l}\}$. Consequently, as crosslinking progresses causing reduction in $\varepsilon$ toward zero, the entire spectrum of relaxation times shifts uniformly to longer times according to $\tau_{max,S} \propto \varepsilon^{-1/\kappa_S}$. However, the relative distribution among the modes, governed by $\{w_{S,l}, B_{S,l}\}$, remains unchanged. Accordingly, when the relaxation modulus is plotted as a function of $t/\tau_{max,S}$, data obtained at different $\varepsilon$ will collapse onto a single master curve characterized by the function $\Phi(x)$. Importantly Eq. (21) generalizes the minimal model by incorporating multiple relaxation times originating from heterogeneous microstructures. The parameters have a clear physical meaning, wherein $w_{S,l}$ represents the fraction of material relaxing through the $l$th mode while $B_{S,l}$ describes the rate of that mode relative to the characteristic timescale $\tau_S \varepsilon^{-1/\kappa_S}$. From a practical standpoint, the value of $N$ and the parameters $\{w_{S,l}, B_{S,l}\}$ can be obtained by standard fitting algorithms developed for Prony series for the given experimental relaxation modulus data. Overall, Eq. (21) provides a general, physically consistent description of relaxation modulus for the complete range of the critical exponent $0 < n < 1$ as the critical gel state is approached from the pre-gel side.

In linear viscoelasticity, the relationship between the stress relaxation modulus $G(t)$ and the continuous relaxation spectrum $H(\tau)$ is given by:[67]

$$G(t,\varepsilon) = G_e + \int_{-\infty}^{\infty} H(\tau,\varepsilon)\exp(-t/\tau)d\ln\tau. \tag{22}$$

For a simple power-law decay, $G(t) = At^{-a}$, the corresponding relaxation spectrum can be analytically computed to be $H(\tau) = A\tau^{-a}/\Gamma(a)$. Consequently, if $G(t,\varepsilon)$ is expressed as a sum of power law terms as shown in Eq. (21), for every term of the form $At^{-a}$ there exists a term $A\tau^{-a}/\Gamma(a)$ in the continuous relaxation spectrum. Consequently, using Eqs. (22) and (21), we obtain the full series expression of the general form of the continuous relaxation spectrum for the pre-gel:

$$H(\tau,\varepsilon) = S \sum_{k=0}^{\infty} \left[\sum_{l=1}^{N} w_{S,l} B_{S,l}^k \right] \frac{(-1)^k \varepsilon^k}{k!\, \Gamma(n-k\kappa_S)} \tau^{-(n-k\kappa_S)}. \tag{23}$$

Furthermore, with knowledge of the general form of the relaxation modulus (Eq. (21)), we also obtain the dynamic moduli using a Fourier transform relation given by:[67]

$$G^*(\omega,\varepsilon) = G'(\omega,\varepsilon) + iG''(\omega,\varepsilon) = i\omega \int_0^{\infty} G(t,\varepsilon) e^{-i\omega t} dt, \tag{24}$$

where $G^*(\omega,\varepsilon)$ is the complex modulus and the storage modulus ($G'(\omega,\varepsilon)$) and the loss modulus ($G''(\omega,\varepsilon)$) are its corresponding real and imaginary parts. Substituting Eq. (21) into Eq. (24) and by performing some manipulation leads to:

$$G^*(\omega,\varepsilon) = iS\omega \sum_{l=1}^{N} w_{S,l} \left( \int_0^{\infty} t^{-n} \left[ \sum_{k=0}^{\infty} \frac{(-1)^k B_{S,l}^k \varepsilon^k t^{k\kappa_S}}{k!} \right] e^{-i\omega t} dt \right), \tag{25}$$

Interchanging the order of summation and integration gives

$$G^*(\omega,\varepsilon) = iS\omega \sum_{k=0}^{\infty} \frac{(-1)^k \varepsilon^k}{k!} \left[ \sum_{l=1}^{N} w_{S,l} B_{S,l}^k \right] \int_0^{\infty} t^{k\kappa_S-n} e^{-i\omega t} dt, \tag{26}$$

The integral can be evaluated using the standard result: $\int_0^{\infty} t^{\alpha} e^{-i\omega t} dt = \Gamma(\alpha+1)(i\omega)^{-\alpha-1}$, which leads to

$$G^*(\omega, \varepsilon) = S \sum_{k=0}^{\infty} \frac{(-1)^k \varepsilon^k}{k!} \left[\sum_{l=1}^{N} w_{S,l} B_{S,l}^k\right] \Gamma(k\kappa_S - n + 1)(i\omega)^{n-k\kappa_S} \qquad (27)$$

Using the identity $(i\omega)^{n-k\kappa_S} = \omega^{n-k\kappa_S} \left[\cos\left(\frac{(n-k\kappa_S)\pi}{2}\right) + i\sin\left(\frac{(n-k\kappa_S)\pi}{2}\right)\right]$, the storage and loss moduli are obtained as

$$G'(\omega, \varepsilon) = S \sum_{k=0}^{\infty} \frac{(-1)^k \varepsilon^k}{k!} \left[\sum_{l=1}^{N} w_{S,l} B_{S,l}^k\right] \Gamma(k\kappa_S - n + 1)\omega^{n-k\kappa_S}\cos\left(\frac{(n-k\kappa_S)\pi}{2}\right) \qquad (28)$$

$$G''(\omega, \varepsilon) = S \sum_{k=0}^{\infty} \frac{(-1)^k \varepsilon^k}{k!} \left[\sum_{l=1}^{N} w_{S,l} B_{S,l}^k\right] \Gamma(k\kappa_S - n + 1)\omega^{n-k\kappa_S}\sin\left(\frac{(n-k\kappa_S)\pi}{2}\right) \qquad (29)$$

The above expressions provide the general and complete frequency-dependent expression of storage and loss modulus of the pre-gel state at any distance $\varepsilon$ from the critical gel point. At the critical gel state ($\varepsilon = 0$), only the $k = 0$ term survives in the series of Eqs. (28) and (29) and if we incorporate $\sum_{l=1}^{N} w_l B_{S,l}^k = 1$, we obtain Eq. (2) confirming the power-law behavior of dynamic moduli at the critical gel state.

Furthermore, in order to examine how the dynamic moduli evolve as the material approaches the critical gel state, we differentiate Eqs. (28) and (29) with respect to $\varepsilon$ leading to:

$$\frac{\partial G'(\omega, \varepsilon)}{\partial \varepsilon} = S \sum_{k=1}^{\infty} \frac{(-1)^k \varepsilon^{k-1}}{(k-1)!} \left[\sum_{l=1}^{N} w_{S,l} B_{S,l}^k\right] \Gamma(k\kappa_S - n + 1)\omega^{n-k\kappa_S}\cos\left(\frac{(n-k\kappa_S)\pi}{2}\right) \qquad (30)$$

$$\frac{\partial G''(\omega, \varepsilon)}{\partial \varepsilon} = S \sum_{k=1}^{\infty} \frac{(-1)^k \varepsilon^{k-1}}{(k-1)!} \left[\sum_{l=1}^{N} w_{S,l} B_{S,l}^k\right] \Gamma(k\kappa_S - n + 1)\omega^{n-k\kappa_S}\sin\left(\frac{(n-k\kappa_S)\pi}{2}\right) \qquad (31)$$

In the limit $\varepsilon \to 0$, only the $k = 1$ term survives, and if we use $B_s = \sum_{l=1}^{N} w_l B_{S,l}^k$, we obtain:

$$\left(\frac{\partial G'(\omega, \varepsilon)}{\partial \varepsilon}\right)_{\varepsilon \to 0} = -SB_s\Gamma(\kappa_S - n + 1)\, \omega^{n-\kappa_S}\cos\left(\frac{(n-\kappa_S)\pi}{2}\right) \qquad (32)$$

$$\left(\frac{\partial G''(\omega, \varepsilon)}{\partial \varepsilon}\right)_{\varepsilon \to 0} = -SB_s\Gamma(\kappa_S - n + 1)\, \omega^{n-\kappa_S}\sin\left(\frac{(n-\kappa_S)\pi}{2}\right) \qquad (33)$$

Dividing the above expressions by the storage and loss modulus associated with the critical gel state (Eq. (2)), leads to the logarithmic derivative given by:

$$\left(\frac{\partial \ln G'(\omega, \varepsilon)}{\partial \varepsilon}\right)_{\varepsilon \to 0} = -B_s \frac{\Gamma(\kappa_S - n + 1)}{\Gamma(1 - n)} \omega^{-\kappa_S} \frac{\cos((n - \kappa_S)\pi/2)}{\cos(n\pi/2)} \quad (34)$$

$$\left(\frac{\partial \ln G''(\omega, \varepsilon)}{\partial \varepsilon}\right)_{\varepsilon \to 0} = -B_s \frac{\Gamma(\kappa_S - n + 1)}{\Gamma(1 - n)} \omega^{-\kappa_S} \frac{\sin((n - \kappa_S)\pi/2)}{\sin(n\pi/2)} \quad (35)$$

As expected, both the logarithmic derivatives scale as $\omega^{-\kappa_S}$, consistent with experimental observations across a wide range of gelling systems.[1, 18, 21-22, 24-30] However, for the first time, we obtain the complete and most general form of these derivatives, which remarkably leads to an analytical expression for the parameter $C_{S \to G}$ introduced in Eq. (13) given by:

$$C_{S \to G} = \frac{\left(\frac{\partial \ln G'(\omega, \varepsilon)}{\partial \varepsilon}\right)_{\varepsilon \to 0}}{\left(\frac{\partial \ln G''(\omega, \varepsilon)}{\partial \varepsilon}\right)_{\varepsilon \to 0}} = \left(\frac{\partial \ln G'(\omega, \varepsilon)}{\partial \ln G''(\omega, \varepsilon)}\right)_{\varepsilon \to 0} = \cot\left(\frac{(n - \kappa_S)\pi}{2}\right) \tan\left(\frac{n\pi}{2}\right)$$

$$\left(\frac{\partial G''}{\partial G'}\right)_{\varepsilon \to 0} = \tan\left(\frac{(n - \kappa_S)\pi}{2}\right)$$

(36)

Eq. (36) establishes an exact relationship between the experimentally measurable parameter $C_{S \to G}$ and the fundamental exponents $n$ and $\kappa_S$. Importantly, $C_{S \to G}$ is independent of frequency $\omega$, the mode coefficients $\{B_{S,l}\}$, and the weight distribution $\{w_{S,l}\}$, and depends solely on the exponents $n$ and $\kappa_S$. Interestingly, while the experimental data suggests $C_{S \to G}$ to be around 2, as a first impression Eq. (36) does not inherently ensure a constant value at first inspection. Furthermore, as mentioned in Eq. (36), $C_{S \to G}$ can equivalently be expressed as $C_{S \to G} = \left(\frac{\partial \ln G'(\omega, \varepsilon)}{\partial \ln G''(\omega, \varepsilon)}\right)_{\varepsilon \to 0}$, which places $C_{S \to G}$ to be a measure of relative change in the logarithm of the storage modulus with respect to that of the loss modulus as the critical gel state is approached. Since at the critical state, $\tan \delta = \tan(n\pi/2)$, Eq. (36) also suggests that $(\partial G'/\partial G'')_{\varepsilon \to 0} = \cot((n - \kappa_S)\pi/2)$. We shall revisit this important issue later in the paper after developing the relaxation modulus and the dynamic moduli for the post-gel states.

**II.2 Post-gel states**

For the post-gel states, we follow an analogous development to that for the pre-gel states. However, there is a fundamental difference that the relaxation modulus for the post-gel states approaches a finite equilibrium modulus $G_e(\varepsilon)$ rather than diminishing at long times. The generic relationship for the relaxation modulus in the post-gel states can be expressed as:

$$G(t,\varepsilon) = St^{-n}\Psi\left(\frac{t}{\tau_{max,G}}\right) + G_e(\varepsilon), \tag{37}$$

where $\tau_{max,G}$ is the characteristic relaxation timescale associated with the relaxable fraction in the post-gel state, given by Eq. (9): $\tau_{max,G} = \tau_G \varepsilon^{-1/\kappa_G}$. The equilibrium modulus follows the power-law scaling given by Eq. (8) : $G_e(\varepsilon) = G_0 \varepsilon^z$. The cutoff function $\Psi(x)$ must satisfy conditions based on the constraints in Table 1 for the post-gel state:

$$\Psi(x) = \begin{cases} 1, & x \to 0 \\ 0, & x \to \infty \end{cases} \tag{38}$$

where the scaling variable is defined as $x = \dfrac{t}{\tau_{max,G}} = \dfrac{t}{\tau_G}\varepsilon^{1/\kappa_G}$.

The experimental observation regarding the logarithmic derivative of dynamic moduli on frequency as $\varepsilon \to 0$ from the post-gel side, given by Eq. (14), can be transformed to obtain a constraint on the relaxation modulus. Consequently, for the relaxable part of the modulus, $G_r(t,\varepsilon) = G(t,\varepsilon) - G_e(\varepsilon)$, we have:

$$\left(\frac{\partial \ln G_r(t,\varepsilon)}{\partial \varepsilon}\right)_{\varepsilon \to 0} = B_g t^{\kappa_G} \tag{39}$$

It is important to note the sign difference compared to the pre-gel case (Eq. (17)), which reflects that $G_r(t,\varepsilon)$ decreases as ε increases due to network consolidation in the post-gel state. Taking the derivative of the logarithm of $G_r(t,\varepsilon)$ from Eq. (37) and performing further manipulation gives:

$$\frac{\partial \ln G_r(t,\varepsilon)}{\partial \varepsilon} = \frac{x}{\kappa_G \varepsilon}\frac{d\ln \Psi(x)}{dx}. \tag{40}$$

In the limit $x \to 0$ (that is equivalent to $\varepsilon \to 0$), in order to recover Eq. (39), the function $\Psi(x)$ must satisfy:

$$\lim_{x \to 0}\left\{\frac{d\ln \Psi(x)}{dx}\right\} = \kappa_G B_g \tau_G^{\kappa_G} x^{\kappa_G - 1} \tag{41}$$

In general, if we consider $\dfrac{d\ln\Psi(x)}{dx} = g(x)$, with the limit as $x \to 0$ given by Eq. (41) and impose $\Psi(x) \to 0$ for $x \to \infty$, we get:

$$\begin{aligned}\Psi(x) &= \exp\left[\int_x^\infty g(u)du\right], \\ u &\to 0, g(u) = \kappa_G B_g \tau_G^{\kappa_G} u^{\kappa_G - 1} \\ x &\to \infty, \int_x^\infty g(u)du = 0\end{aligned} \tag{42}$$

The two different integration forms, Eq. (42) for the post-gel states (integration from $x$ to $\infty$) and Eq. (20) for the pre-gel states (integration from 0 to $x$) arise directly from the opposite signs of the experimentally observed logarithmic derivative of the relaxation modulus with respect to $\varepsilon$ on the two sides of the critical gel point. Furthermore, Eq. (42) naturally ensures $\Psi(x) \to 1$ as $x \to 0$ and $\Psi(x) \to 0$ as $x \to \infty$. Therefore, any function $g(u)$ that satisfies the expression and constraints mentioned in Eq.(42) is an acceptable form of the cutoff function $\Psi(x)$ to be incorporated in the expression of the relaxation modulus for the post-gel states.

Now, similar to the pre-gel states, if we assume that the asymptotic form $g(u) = v_G B_g \tau_G^{\kappa_G} u^{\kappa_G - 1}$ applies over the entire range of $u$ (from 0 to $\infty$), we get:

$$\Psi(x) = \exp\left[\int_x^\infty \kappa_G B_g \tau_G^{\kappa_G} u^{\kappa_G - 1} du\right] = \exp\left(+B_g \tau_G^{\kappa_G} x^{\kappa_G}\right), \tag{43}$$

Substituting Eq. (43) into Eq. (37), leads to the corresponding expression for the relaxation modulus given by:

$$G(t, \varepsilon) = S t^{-n} \exp\left(+B_g \tau_G^{\kappa_G} x^{\kappa_G}\right) + G_e(\varepsilon) \tag{44}$$

However, as discussed by Suman and coworkers,[22] in this form, the exponential term grows without bound as $t \to \infty$, which violates a constraint that $\left(\frac{\partial G(t,\varepsilon)}{\partial t}\right)_\varepsilon \leq 0$. To obtain a physically acceptable form, $g(u)$ needs to be modified in such a way that it decays sufficiently rapidly at large $u$ to ensure the integral in Eq. (42) vanishes as $x \to \infty$ and $G(t, \varepsilon)$ does not increase with $t$ for a given $\varepsilon$.

One proposal, due to Suman and coworkers,[22] is to truncate the series expansion. The corresponding expression (Eq. (11)) can be rewritten for the relaxable part of the modulus given by:

$$G_r(t, \varepsilon) = S t^{-n} \left(\sum_{m=0}^{\lfloor n/\kappa_G \rfloor} \frac{(B_g \varepsilon)^m t^{m\kappa_G}}{m!}\right), \tag{11}$$

where the floor function $\lfloor n/\kappa_G \rfloor$ represents the largest integer smaller than or equal to $n/\kappa_G$. Interestingly, this form naturally satisfies all the constraints outlined in Table 1 for the post-gel state. However, its applicability depends critically on the relative magnitudes of $n$ and $\kappa_G$. When $n \geq \kappa_G$, the series contains multiple terms ($\lfloor n/\kappa_G \rfloor \geq 1$), providing a rich description of the relaxation dynamics. For example, if $n = 0.65$ and $\kappa_G = 0.2$, then $\lfloor 0.65/0.2 \rfloor = 3$, and the series includes terms with $m = 0, 1, 2, 3$. This captures the relaxation dynamics in the post-gel state well. When $n < \kappa_G$, the floor function leads to $\lfloor n/\kappa_G \rfloor = 0$, and the series in Eq. (11) contains only the $m = 0$ term. We believe that the regime for which $n < \kappa_G$ is physically inadmissible, and if at all it exists in reality, the post-gel relaxation dynamics would be governed by fundamentally different physics. A detailed discussion of the constraint $n \geq \kappa_G$ and its physical implications is provided in Section II.4. We, therefore, note that the framework developed in this section, including the truncated series representation and all subsequent expressions for dynamic moduli and their derivatives mentioned below, is applicable only when $n \geq \kappa_G$.

For systems satisfying $n \geq \kappa_G$, while Eq. (11) represents a general obvious analytical form for the post-gel relaxation modulus, it is important to acknowledge that other functional forms satisfying the constraints in Eq. (42) may exist. Following the framework developed for the pre-gel state, the general form for $G_r(t, \varepsilon)$ can also be described in terms of a multi-mode exponential representation given by:

$$G_r(t, \varepsilon) = S t^{-n} \sum_{l=1}^{M} w_{G,l} \left[ \sum_{m=0}^{\lfloor n/\kappa_G \rfloor} \frac{(B_{G,l}\varepsilon)^m}{m!} t^{m\kappa_G} \right], \quad (45)$$

where $M$ represents the number of modes, $w_{G,l} > 0$ are weight factors satisfying the normalization constraint $\sum_{l=1}^{M} w_{G,l} = 1$, and $B_{G,l} > 0$ are mode-specific coefficients. After some manipulation of Eq. (45) the generic form for the relaxation modulus in the post-gel state is given by:

$$G(t, \varepsilon) = G_r(t, \varepsilon) + G_e(\varepsilon) = S \left[ \sum_{m=0}^{\lfloor n/\kappa_G \rfloor} \left[ \sum_{l=1}^{M} w_{G,l} B_{G,l}^m \right] \frac{\varepsilon^m}{m!} t^{m\kappa_G - n} \right] + G_e(\varepsilon). \quad (46)$$

For $M = 1$, Eq. (46) reduces to the single-mode form given by Eq. (11). A critical constraint for any acceptable form of relaxation modulus is: $(\partial G/ \partial t)_\varepsilon \leq 0$, which ensures that the stress relaxation proceeds monotonically. Taking the time derivative of Eq. (46):

$$\frac{\partial G_r(t, \varepsilon)}{\partial t} = \frac{\partial G(t, \varepsilon)}{\partial t} = S \sum_{m=0}^{\lfloor n/\kappa_G \rfloor} \left[ \sum_{l=1}^{M} w_{G,l} B_{G,l}^m \right] \frac{\varepsilon^m}{m!} (m\kappa_G - n) t^{m\kappa_G - n - 1}, \quad (47)$$

For each term in the double sum, the coefficient $(m\kappa_G - n)$ determines the sign. For all $m \leq \lfloor n/\kappa_G \rfloor$, we have $m\kappa_G \leq \lfloor n/\kappa_G \rfloor \kappa_G < n$. Therefore, $(m\kappa_G - n) < 0$ for all $m$ in the summation range, guarantees that $\partial G_r/ \partial t < 0$ for all $t > 0$ and $\varepsilon > 0$ satisfying the monotonicity constraint without imposing any additional restrictions on $\{w_{G,l}, B_{G,l}\}$ beyond their positivity and the normalization of weights. The multi-mode representation in Eq. (46) provides a general framework for describing the relaxation modulus in the post-gel states when $n \geq \kappa_G$. Similar to the pre-gel case, this formulation can be perceived as a generalized Prony series, capable of capturing various simultaneous relaxation processes arising from distributed network architectures. Furthermore, similar to the pre-gel states, time-cure superposition in the post-gel state requires that changes in $\varepsilon$ to affect only the characteristic timescale $\tau_{max,G}$ without altering the shape of the relaxation time spectrum. The relaxable component should be expressible as $G_r(t,\varepsilon) = St^{-n}\Psi(t/\tau_{max,G})$, where the cutoff function $\Psi$ depends on $\varepsilon$ only through the ratio $t/\tau_{max,G}$. Similar to that for the pre-gel states, the proposed Eq. (46) confirms that time-cure superposition is satisfied. The time-cure superposition is validated by the present theoretical treatment, which can be extended to any linear viscoelastic parameter, $G(t,\varepsilon)$, $G'(\omega,\varepsilon)$, $G'(\omega,\varepsilon)$ and tanδ as has been reported experimentally.[1, 18, 22, 30, 34, 63-65] We can also obtain an expression for the continuous relaxation time spectrum by incorporating Eq. (46) into Eq. (22). By performing term-by-term conversion similar to that was done for the pre-gel states, the general expression for $H(\tau,\varepsilon)$ for the post-gel states is given by:

$$H(\tau,\varepsilon) = S \sum_{m=0}^{\lfloor n/\kappa_G \rfloor} \left[\sum_{l=1}^{M} w_{G,l} B_{G,l}^m\right] \frac{\varepsilon^m}{m!\, \Gamma(n-m\kappa_G)} \tau^{-(n-m\kappa_G)}. \tag{48}$$

The dynamic moduli for Eq. (46) are obtained through the Fourier transform mentioned in Eq. (24). When the relaxation modulus for the post-gel state is expressed as $G(t,\varepsilon) = G_r(t,\varepsilon) + G_e(\varepsilon)$, the corresponding storage and loss modulus result in: $G'(\omega,\varepsilon) = G'_r(\omega,\varepsilon) + G_e(\varepsilon)$ and $G''(\omega,\varepsilon) = G''_r(\omega,\varepsilon)$, where $G'_r$ and $G''_r$ are the contributions from the relaxable components. If we represent $G_r^*(\omega,\varepsilon) = G'_r(\omega,\varepsilon) + iG''_r(\omega,\varepsilon)$, it can be expressed in terms of the Fourier transform of $G_r(t,\varepsilon)$ mentioned in Eq. (46) as:

$$G_r^*(\omega,\varepsilon) = i\omega \int_0^\infty S\left[\sum_{m=0}^{\lfloor n/\kappa_G \rfloor} \frac{\varepsilon^m}{m!}\left[\sum_{l=1}^{M} w_{G,l} B_{G,l}^m\right] t^{m\kappa_G-n}\right] e^{-i\omega t} dt, \tag{49}$$

By carrying out analysis similar to that for the pre-gel state, the complete expression for the dynamic moduli for $n \geq \kappa_G$, in the post gel states, is given by:

$$G'(\omega, \varepsilon) = S \sum_{m=0}^{\lfloor n/\kappa_G \rfloor} \frac{\varepsilon^m}{m!} \left[ \sum_{l=1}^{M} w_{G,l} B_{G,l}^m \right] \Gamma(m\kappa_G - n + 1) \omega^{n-m\kappa_G} \cos\left(\frac{(n - m\kappa_G)\pi}{2}\right) + G_0 \varepsilon^z \quad (50)$$

$$G''(\omega, \varepsilon) = S \sum_{m=0}^{\lfloor n/\kappa_G \rfloor} \frac{\varepsilon^m}{m!} \left[ \sum_{l=1}^{M} w_{G,l} B_{G,l}^m \right] \Gamma(m\kappa_G - n + 1) \omega^{n-m\kappa_G} \sin\left(\frac{(n - m\kappa_G)\pi}{2}\right) \quad (51)$$

At the critical gel state ($\varepsilon = 0$), the equilibrium modulus vanishes ($G_e(0) = 0$), and only the $m = 0$ term survives in the series resulting in Eq. (2). To assess the evolution of dynamic moduli as the material moves away from the critical gel state into the post-gel regime (or the other way around), we differentiate Eqs. (50) and (51) with respect to $\varepsilon$ leading to:

$$\frac{\partial G'(\omega, \varepsilon)}{\partial \varepsilon} = S \sum_{m=1}^{\lfloor n/\kappa_G \rfloor} \frac{\varepsilon^{m-1}}{(m-1)!} \left[ \sum_{l=1}^{M} w_{G,l} B_{G,l}^m \right] \Gamma(m\kappa_G - n + 1) \omega^{n-m\kappa_G} \cos\left(\frac{(n - m\kappa_G)\pi}{2}\right) + G_0 z \varepsilon^{z-1} \quad (52)$$

$$\frac{\partial G''(\omega, \varepsilon)}{\partial \varepsilon} = S \sum_{m=1}^{\lfloor n/\kappa_G \rfloor} \frac{\varepsilon^{m-1}}{(m-1)!} \left[ \sum_{l=1}^{M} w_{G,l} B_{G,l}^m \right] \Gamma(m\kappa_G - n + 1) \omega^{n-m\kappa_G} \sin\left(\frac{(n - m\kappa_G)\pi}{2}\right) \quad (53)$$

For systems with $n \geq \kappa_G$, in the limit $\varepsilon \to 0$, only the $m = 1$ term from the series survives, and the equilibrium modulus term vanishes, resulting in:

$$\left(\frac{\partial G'(\omega, \varepsilon)}{\partial \varepsilon}\right)_{\varepsilon \to 0} = S B_g \Gamma(\kappa_G - n + 1) \omega^{n-\kappa_G} \cos\left(\frac{(n - \kappa_G)\pi}{2}\right) \quad (54)$$

$$\left(\frac{\partial G''(\omega, \varepsilon)}{\partial \varepsilon}\right)_{\varepsilon \to 0} = S B_g \Gamma(\kappa_G - n + 1) \omega^{n-\kappa_G} \sin\left(\frac{(n - \kappa_G)\pi}{2}\right) \quad (55)$$

where $B_g = \sum_{l=1}^{N} w_{G,l} B_{G,l}^m$. It is important to note the sign difference compared to the pre-gel case (Eqs. (32) and (33)). It essentially originates because an increase in $\varepsilon$ precisely means the opposite on the either side of the critical gel state. In the pre-gel states, decrease in $\varepsilon$ means increase in degree of crosslinking (the derivatives are negative). In case of the post-gel states, on the other hand, as $\varepsilon$ increases, the network consolidates due to an increase in the degree of crosslinking (the derivatives are positive).

The logarithmic derivatives of $G'(\omega, \varepsilon)$ and $G''(\omega, \varepsilon)$ are obtained by dividing (Eqs. (54) and (55)) by the moduli at the critical gel state given by Eq. (2) given by:

$$\left(\frac{\partial \ln G'(\omega, \varepsilon)}{\partial \varepsilon}\right)_{\varepsilon \to 0} = B_g \frac{\Gamma(\kappa_G - n + 1)}{\Gamma(1 - n)} \omega^{-\kappa_G} \frac{\cos\left(\frac{(n - \kappa_G)\pi}{2}\right)}{\cos\left(\frac{n\pi}{2}\right)} \tag{56}$$

$$\left(\frac{\partial \ln G''(\omega, \varepsilon)}{\partial \varepsilon}\right)_{\varepsilon \to 0} = B_g \frac{\Gamma(\kappa_G - n + 1)}{\Gamma(1 - n)} \omega^{-\kappa_G} \frac{\sin\left(\frac{(n - \kappa_G)\pi}{2}\right)}{\sin\left(\frac{n\pi}{2}\right)} \tag{57}$$

Both logarithmic derivatives exhibit the characteristic frequency scaling $\omega^{-\kappa_G}$, in accordance with the experimental observations on gel-sol transitions approached from the post-gel side.[27, 31] Finally, for $n \geq \kappa_G$, the ratio of the logarithmic derivatives yields:

$$\begin{aligned} C_{G \to S} = \left(\frac{\partial \ln G'(\omega, \varepsilon)}{\partial \ln G''(\omega, \varepsilon)}\right)_{\varepsilon \to 0} &= \cot\left(\frac{(n - \kappa_G)\pi}{2}\right) \tan\left(\frac{n\pi}{2}\right) \\ \left(\frac{\partial G''}{\partial G'}\right)_{\varepsilon \to 0} &= \tan\left(\frac{(n - \kappa_G)\pi}{2}\right) \end{aligned} \tag{58}$$

Remarkably, this expression is identical in form to that obtained for the pre-gel state (Eq. (36)), with $\kappa_S$ replaced by $\kappa_G$. Similar to that mentioned for the pre-gel states, from the post-gel side, we get: $(\partial G'/\partial G'')_{\varepsilon \to 0} = \cot((n - \kappa_G)\pi/2)$. Furthermore, Eq. (58) clearly suggests that one cannot a priory assign any specific value to $C_{G \to S}$, except that $C_{G \to S}$ is necessarily positive.

## II.3 Continuity Constraints at the Critical Gel Point

A fundamental requirement for any physically consistent description of the sol-gel transition is that the viscoelastic properties must vary continuously as the material passes through the critical gel point. While the linear viscoelastic properties, such as relaxation modulus and dynamic moduli, change their functional form at $p = p_c$, the respective forms in the pre-gel state and the post-gel state, do approach the critical gel representation in the limiting case of $p \to p_c$ from either side. However, it is not merely important that the properties themselves are continuous as sol transforms to gel, or vice versa; it is also important that the derivatives of these properties with respect to the degree of crosslinking must be continuous across the sol-gel transition. This requirement of continuity of derivatives imposes stringent constraints on the scaling exponents and material parameters, providing a bridge between the pre-gel and post-gel expressions developed in Sections II.1 and II.2. More importantly, these constraints provide profound insights into the dynamics of the sol-gel transition.

In the introduction section, we define the distance from the gel point as $\varepsilon = |p - p_c|$. Consequently, for the pre-gel regime ($p < p_c$), we have $\varepsilon = p_c - p$, while for the post-gel regime ($p > p_c$), we have $\varepsilon = p - p_c$. Accordingly, derivatives with respect to $p$ transform as:

$$\frac{\partial}{\partial p} = \begin{cases} -\dfrac{\partial}{\partial \varepsilon}, & p < p_c \text{ (pre-gel)} \\ +\dfrac{\partial}{\partial \varepsilon}, & p > p_c \text{ (post-gel)} \end{cases} \tag{59}$$

This sign difference is crucial and reflects the fact that increasing $p$ corresponds to decreasing $\varepsilon$ in the pre-gel state (denoting approach to the critical gel state) but amounts to increasing $\varepsilon$ in the post-gel state (implying moving away from the critical gel state).

While one can consider the derivative of any property to enforce continuity, dynamic moduli are the most convenient parameters, as their evolution across the sol-gel transition can be explicitly measured experimentally. The storage and loss moduli can be converted from $\varepsilon$ representation to $p$ and $p_c$ representation, by replacing the former by $p_c - p$ in Eqs. (28) and (29) (the pre-gel state) while by $p - p_c$ in Eqs. (50) and (51) (the post gel state). Using $\partial / \partial p = - \partial / \partial \varepsilon$ in the pre-gel regime, the derivatives represented in Eqs. (32) and (33) with respect to $p$ become:

$$\left(\frac{\partial G'(\omega, p)}{\partial p}\right)_{p \to p_c^-} = +SB_s\Gamma(\kappa_S - n + 1)\,\omega^{n-\kappa_S}\cos\left(\frac{(n - \kappa_S)\pi}{2}\right) \tag{32}$$

$$\left(\frac{\partial G''(\omega, p)}{\partial p}\right)_{p \to p_c^-} = +SB_s\Gamma(\kappa_S - n + 1)\,\omega^{n-\kappa_S}\sin\left(\frac{(n - \kappa_S)\pi}{2}\right) \tag{33}$$

On the other hand, since $\partial / \partial p = + \partial / \partial \varepsilon$ in the post-gel regime, the derivatives mentioned in Eqs. (54) and (55) with respect to $p$ are:

$$\left(\frac{\partial G'(\omega, p)}{\partial p}\right)_{p \to p_c^+} = +SB_g\Gamma(\kappa_G - n + 1)\omega^{n-\kappa_G}\cos\left(\frac{(n - \kappa_G)\pi}{2}\right) \tag{54}$$

$$\left(\frac{\partial G''(\omega, p)}{\partial p}\right)_{p \to p_c^+} = +SB_g\Gamma(\kappa_G - n + 1)\omega^{n-\kappa_G}\sin\left(\frac{(n - \kappa_G)\pi}{2}\right) \tag{55}$$

For a physically consistent description of the sol-gel transition, the derivatives of the dynamic moduli with respect to $p$ must be continuous at the critical gel point. Requiring continuity of the derivative of the storage modulus:

$$\left(\frac{\partial G'(\omega, p)}{\partial p}\right)_{p \to p_c^-} = \left(\frac{\partial G'(\omega, p)}{\partial p}\right)_{p \to p_c^+} \text{ and } \left(\frac{\partial G''(\omega, p)}{\partial p}\right)_{p \to p_c^-} = \left(\frac{\partial G''(\omega, p)}{\partial p}\right)_{p \to p_c^+} \tag{60}$$

If we impose these conditions on Eqs. (32) and (33) and Eqs. (54) and (55), especially for this equality to hold for all frequencies $\omega$, the following two conditions must be met:

1. Frequency exponents must match, implying:

$$\kappa_S = \kappa_G \equiv \kappa. \tag{61}$$

2. And the corresponding coefficients must match, leading to:

$$B_S = \sum_{l=1}^{N} w_{S,l} B_{S,l} = \sum_{l=1}^{M} w_{G,l} B_{G,l} = B_G. \tag{62}$$

An alternative and equivalent perspective on this continuity requirement can be obtained by examining the equality of $C_{S \to G} = \left(\frac{\partial G'}{\partial G''}\right)_{p \to p_c^-} = \left(\frac{\partial G'}{\partial G''}\right)_{p \to p_c^+} = C_{G \to S}$ We have discussed this aspect in further detail in Appendix A. Interestingly, equality of Eqs. (36) and (58)) also results in Eq. (61). It should be noted that Eqs. (61) and (62) requiring the mentioned equalities are not assumptions or empirical observations but necessary conditions for physical consistency derived from the general expressions of the relaxation modulus. Therefore, in principle, any material undergoing a continuous sol-gel transition and adhering to the constraints mentioned above must satisfy Eqs. (61) and (62).

The condition $\kappa_S = \kappa_G = \kappa$ is a very significant constraint and is model-agnostic, as discussed below. As mentioned in the introduction section $\kappa_S = \kappa_G$ enforces symmetry in the evolution of the relaxation times on both sides of the gel point. Therefore, following Eqs. (5) and (9) the longest relaxation time in the pre-gel and post-gel states should respectively diverge as $\tau_{max,S} \propto \varepsilon^{-1/\kappa}$ and $\tau_{max,G} \propto \varepsilon^{-1/\kappa}$, reflecting a dynamical balance in how the system approaches and departs from the critical gel state. Furthermore, equality $\kappa_S = \kappa_G$, according to Eq. (12), necessarily leads to validation of the hyperscaling relationship $(n = z/(z+s))$. This equality also facilitates a consistency check for the entire theoretical framework: if independent measurements of $n$, $s$, and $z$ satisfy the hyperscaling relationship $(n = z/(z+s))$, then the system exhibits symmetric gelation dynamics, and the formulations developed in Sections II.1 and II.2 provide a complete and consistent description of the sol-gel transition. Conversely, if this hyperscaling relationship holds, then $\kappa_S = \kappa_G$ also follows automatically.

It is important to note that the origin of the prescribed symmetry condition $\kappa_S = \kappa_G = \kappa$ is in the specific dynamical constraints imposed on the system as it approaches the critical state. The analysis suggests that the logarithmic derivatives of the relaxation modulus (Eqs. (17) and (39)) are the mathematical drivers of this symmetry. Specifically, these derivatives dictate the asymptotic behavior of the cutoff functions $\Phi(x)$ and $\Psi(x)$, which in turn enforce the specific frequency dependence of the logarithmic derivatives of dynamic moduli as $\omega^{-\kappa_S}$ (Eqs. (13) or (34) and (35)) and $\omega^{-\kappa_G}$ (Eqs. (14) or (56) and (57)) respectively. It is only because the system obeys these specific logarithmic scaling constraints on both sides of the transition, as experimentally observed for a large number of gel-forming systems, that the requirement of continuity at the gel point forces the relaxation exponents to match. If these constraints were removed or relaxed, which is yet to be explicitly demonstrated for any gel-forming system, the frequency dependence of the evolving moduli would likely differ, and physical continuity at $p_c$ would not necessitate $\kappa_S = \kappa_G$. Therefore, the experimentally validated scaling of the logarithmic derivatives is at the heart of the symmetric evolution of viscoelasticity across the sol-gel transition. Consequently, the validity of the hyperscaling relation ($n = z/(z+s)$) serves as a rigorous consistency check for the nature of the transition, symmetric or otherwise. Hence, deviations from the equality $\kappa_S = \kappa_G$ (and by extension the hyperscaling relation) carry significant diagnostic value. In experimental practice, such deviations may signal measurement artifacts, particularly errors in localizing the critical gel point ($p_c$, $T_c$ or $t_c$) or inaccuracies in determining the longest relaxation times. However, if experimental error is ruled out, $\kappa_S \neq \kappa_G$ may indicate that the fundamental physics of the system lies beyond the framework of continuous equilibrium critical phenomena. Such asymmetry, therefore, may imply that the transition is discontinuous (first order), governed by non-equilibrium dynamics such as aging, progress in the network formation over a single oscillation cycle (mutation numbers are larger than that prescribed), anisotropic nature of the resultant gel, or that the structural arrest mechanism differs from the percolation theory. Such deviations may lead to violation of the assumption of continuity in the derivatives of the dynamic moduli.

**II.4 The Post-Gel Framework for $n < \kappa$**

In this work, we have developed general expressions for the stress-relaxation modulus and dynamic moduli in both pre-gel and post-gel states. As noted in Section II.2, for the applicability of the theoretical framework, the critical exponent $n$ and the relaxation scaling exponent $\kappa$ must satisfy $n \geq \kappa$. The constraint emerges from the mathematical structure of the post-gel relaxation modulus (Eq. (46) and also for Eq. (11)), wherein the relaxable component is expressed as a truncated series that arises from the requirement that the relaxation modulus must decrease monotonically with time. For $n \geq \kappa_G$, the series contains multiple terms ($m = \lfloor n/\kappa_G \rfloor \geq 1$), wherein the $m = 1$ term, in particular, governs the leading-order behavior as $\varepsilon \to 0$. However, when $n < \kappa_G$, we have $\lfloor n/\kappa_G \rfloor = 0$, and the series gets reduced to only the $m = 0$ term resulting in $G_r(t, \varepsilon) = St^{-n}$. Paradoxically this expression is independent of $\varepsilon$, indicating the relaxable component possesses the identical form as that of the critical gel regardless of the distance from the gel point. Consequently, all evolution in the post-gel state would be carried entirely by the equilibrium modulus $G_e(\varepsilon) = G_0 \varepsilon^z$.

The absence of the $m = 1$ term when $n < \kappa$ has profound consequences for the scaling behavior near the critical gel state. The constraint on the time-domain relaxation modulus, $\left(\frac{\partial \ln r}{\partial \varepsilon}\right)_{\varepsilon \to 0} \sim t^{\kappa_G}$ given by Eq. (39) cannot be satisfied as $G_r(t, \varepsilon) = St^{-n}$ is independent of $\varepsilon$. Accordingly, the frequency-Domain Scaling (Eqs. (52) and (53)) given by $\left(\frac{\partial \ln G^*}{\partial \varepsilon}\right)_{\varepsilon \to 0} \sim \omega^{-\kappa_G}$ also becomes inapplicable. The most striking consequence of the above-mentioned inapplicability for $n < \kappa_G$ is the failure of the continuity requirement at the critical gel point discussed in the previous section. For a physically consistent sol-gel transition, the derivatives of the dynamic moduli with respect to the degree of crosslinking $p$ must be continuous as the material passes through the critical point $p_c$. From the pre-gel side the derivatives (Eqs. (32) and (33)) are always finite for all $n \in (0,1)$. However, from the post-gel side, when $n < \kappa_G$, since the $m = 1$ term is absent from the truncated series, the derivative of storage modulus is dominated by the equilibrium modulus contribution: $\left(\frac{\partial G'}{\partial \varepsilon}\right)_{\varepsilon \to 0} = G_0 z \varepsilon^{z-1}$. From the scaling arguments mentioned in the Appendix A, in the post-gel region, we have $\kappa_G = n/z$ (Eq. (A10)) notwithstanding how large or small $n$ is with respect to $\kappa_G$. However, when $n < \kappa_G$, Eq. A10 leads to $z < 1$, which implies $\left(\frac{\partial G'}{\partial \varepsilon}\right)_{\varepsilon \to 0} \to \infty$. Accordingly, the divergence of post-gel derivative of the storage modulus makes it impossible to match the finite pre-gel derivative of the same. Consequently, the continuity requirement cannot be satisfied for $n < \kappa_G$. Furthermore, de Gennes proposed that if $\xi$ is the correlation length, it diverges as system approaches the critical gel state as: $\xi \sim \varepsilon^{-\nu}$, with $\nu$ being a positive constant. For a 3-dimensional network, he proposed $z = 1 + \nu$, clearly suggesting $z$ cannot have a value smaller than unity.[68]

Interestingly the experimental literature on sol-gel transition has reported $\kappa \approx 0.2$ as a seemingly universal value (practically values between 0.1 and 0.3 are reported), for a very broad class of gel-forming materials, including chemically crosslinking polymers as well as spontaneous as well as thermos-reversible aggregating colloidal suspensions. However, experimental studies probing systems with small critical exponents ($n < 0.2$) are virtually absent from the literature. This raises a fundamental question: Is $\kappa \approx 0.2$ truly universal? If $\kappa$ is indeed a universal constant independent of $n$, then the present work puts a lower bound of $\kappa$ on $n$. However, if at all any experimental systems were observed with $n \leq \kappa$, it may have alternate gelation mechanisms, which are not governed by percolation theory. On the other hand, such systems could be affected by non-equilibrium dynamics, such as aging (strong time-dependency) or kinetic arrest, which lie outside the scope of equilibrium critical phenomena. The other possibility is that relaxation scaling exponent $\kappa$ may not be universal over the entire range of $n$, but rather a function of the critical exponent: $\kappa = \kappa(n)$, especially in the low $n$ limit, such that $\kappa < n$ is always maintained. In such cases, the framework developed in this work would apply universally. It should be noted that the fractal dimension at the critical state (Eq. (3)) approaches $f_d \to 2.5$ as $n \to 0$, which indicates a very compact, dense structure. The relaxation dynamics of such dense networks, therefore, might indeed be characterized by smaller values of $\kappa$.

## II.5 Consideration of the faster relaxation modes

It is important to note that the theoretical framework developed in this work focuses exclusively on that region of the relaxation spectrum as identified by the Winter-Chambon (WC) criterion. In this region, the self-similar critical gel behavior manifests leading to characteristic signatures of the sol-gel transition given by power-law relaxation, divergence of the longest relaxation time, and growth of the equilibrium modulus in the post-gel state. However, in practice, the experimentally accessible frequency range may also encompass the faster relaxation modes arising from segmental dynamics, entanglement relaxation, or interparticle interactions. Such fast modes constitute a precursor contribution at short times and high frequencies. This aspect has been discussed in detail by Mours and Winter,[21] wherein a complete description of the viscoelastic response across the full frequency range is obtained by superposing the Baumgärtel–Schausberger–Winter (BSW) spectrum,[69] which captures the fast relaxation modes, with the spectrum in the WC regime discussed here. However, it should be noted that the BSW contribution remains unchanged during the process of gelation. This is because the chemical or physical crosslinks do not affect the local segmental and entanglement dynamics. The parameters of the BSW spectrum are therefore needed to be determined independently of the uncrosslinked precursor material.[21] The expressions developed in Sections II.1–II.4 are therefore valid and intended for application in the CW regime, and accordingly, the experimental data analysed in Section III are selected. For more details, readers interested in the combined BSW+WC framework may refer to the comprehensive treatment by Mours and Winter.[28]

## II.6 Issue of parameters identification

Before proceeding to analyse the experimental data, it is important to address a critical question: the identification of parameters that adhere to the mathematical structure of the expressions derived above. For the pre-gel states, in the expressions of $G(t,\varepsilon)$ (Eq. (21)) as well as the expressions of $G'(\omega,\varepsilon)$ (Eq. (28)) and $G''(\omega,\varepsilon)$ (Eq. (29)), the distance parameter $\varepsilon$ and the mode coefficients $B_{S,l}$ appear exclusively through the products $\varepsilon B_{S,l}$ in every term of order $k$. This implies if we perform a rescaling transformation $\varepsilon \to \beta\varepsilon$ and $B_{S,l} \to B_{S,l}/\beta$ for all $l$, with any $\beta > 0$, all the three response functions $G(t,\varepsilon)$, $G'(\omega,\varepsilon)$ and $G''(\omega,\varepsilon)$ remain exactly the same, since the every product $\varepsilon B_{S,l}$ in the sum is unchanged. The conditions $\sum_{l=1}^{N} w_{S,l} = 1$ and $w_{S,l} > 0$ do not impose any restriction on $\beta$, as they constrain only the weights. Consequently, for any solution ($\varepsilon B_{S,l}$) that fits the data well, there exists a continuous one-parameter family of equally valid solutions ($\beta\varepsilon$, $B_{S,l}/\beta$), that lead to identical rheological predictions. Equivalently, in the post-gel regime, the same degeneracy ($\beta\varepsilon$, $B_{G,l}/\beta$) applies for $G(t,\varepsilon)$, $G'(\omega,\varepsilon)$ and $G''(\omega,\varepsilon)$ (respectively for Eqs. (46), (50) and (51) as well. There is indeed an additional term for the post-gel states: $G_e = G_0 \varepsilon^z$. But, in this expression, any change $\varepsilon \to \beta\varepsilon$ gets accommodated as: $G_0 \to G_0/\beta^z$ without affecting the final fitting. One might expect that the physically motivated constraint $\varepsilon < 1$ resolves this degeneracy. However, this constraint does not help, as any solution with $\varepsilon < 1$, one can always choose $\beta \in (0,1)$ to obtain $\beta\varepsilon < \varepsilon < 1$, so the rescaled solution also satisfies the constraint. Similarly, the requirement $\varepsilon \to 0$ at the critical gel state also does not provide any additional benefit, because as $\varepsilon \to 0$, we have $\beta\varepsilon \to 0$ for any finite $\beta$ simultaneously. This discussion, therefore, leads to a profound conclusion that the absolute numerical value of $\varepsilon$ cannot be uniquely determined from rheological data alone, regardless of any normalization convention that one may adopt.

Typically, in the experimental literature, the expression for the reduced distance $\varepsilon$ is based on an independent parameter, such as temperature, time (spontaneous gelation), concentration of a species, pH, etc. If $X$ is the quantifiable value of that critical parameter ($X = T, t, ...$), then $\varepsilon$ is tentatively defined as $\varepsilon = |X - X_C|/X_C$, where $X_C$ is the value of that parameter at the critical state. The validation of the scaling analysis does lead to universal scaling power law exponents, but the proportionality constants multiplying these power laws are non-universal and depend on microscopic details and normalization conventions. It does not matter how $\varepsilon$ is defined beyond $\varepsilon \sim |X - X_C|$ as any constant of proportionality results in the same scaling exponents. On the other hand, the constants of proportionality are never reported, as the physics is carried entirely by the exponents. Therefore, the genuinely physical content of the model, which is independent of the normalization convention, consists of: (i) the critical exponents $n$ and $\kappa$; (ii) the gel strength $S$; and (iii) the ratios $\varepsilon$(state 1)/$\varepsilon$(state 2). The last condition ensures that any proportionality constant cancels identically. In this work, we base all physical conclusions exclusively on these convention-independent quantities. The absolute values of $\varepsilon$ and $B_{S,l}$ (Or $B_{G,l}$) that we report follow the degeneracy ($\beta\varepsilon$, $B_{S,l}/\beta$) and therefore can be rescaled without affecting any physical result.

## III. Results and Discussion

The theoretical framework developed in the previous section provides the general expressions for the relaxation modulus $G(t,\varepsilon)$ and dynamic moduli ($G'(\omega,\varepsilon)$ and $G''(\omega,\varepsilon)$) in the pre-gel and post-gel states. In this section, we assess the validity of these expressions by comparing them against experimental data already published in the literature. We also examine the physical insights that emerge from this analysis. Since this work emphasizes symmetry of the relaxation dynamics on both the sides of the critical gel state, for remaining part of the paper, we use a single value of $\kappa$ for both the pre-gel as well as the post-gel states. With this background, we first consider the relaxation modulus data on both sides of the sol-gel transition. However, the experimental relaxation modulus data in the vicinity of the critical gel state are scarce in the literature. Therefore, we employ the data reported by Winter and coworkers[1, 70-71] on a chemically crosslinked poly(dimethylsiloxane) (PDMS) system that is transformed from the frequency-domain dynamic moduli to the time-domain relaxation modulus. In this system, Mours and Winter[71] arrest the crosslinking reaction in PDMS at various degrees of crosslinking, and the resulting dynamic moduli were measured over a broad range of frequencies using time-temperature superposition. The relaxation modulus at different degrees of crosslinking was subsequently obtained by performing an inverse Fourier transform, leading to a dataset in the time domain that spans the pre-gel and post-gel states.

Figure 1(a) shows fits of Eqs. (21) and (46) with four modes to this experimental dataset by Winter and coworkers[1, 70-71] across all six states simultaneously, wherein a single consistent set of material parameters listed in Table 2 is used. It can be seen that the quality of the fits is excellent for both pre-gel and post-gel states across all the data sets, spanning more than six decades in $t/a_T$ and nearly five decades in $G$. We believe that this simultaneous description of the full transition with a unique parameter set is a profound outcome and suggests strong validation of the theoretical framework developed in this paper. The critical gel state ($\varepsilon = 0$) shown by a dashed black line manifests as a power law $G = St^{-n}$ that is devoid of any characteristic timescale. As $\varepsilon$ increases in the pre-gel state, the relaxation modulus gradually departs from this power law at shorter times. Furthermore, the exponential cutoff imposed by the finite maximum relaxation time $\tau_{max,S} \sim \varepsilon^{-1/\kappa}$ becomes evident through steep terminal decay of the pre-gel relaxation modulus curves. On the other hand, in the post-gel state, the relaxation modulus curves lie above the critical gel line and develop a long-time plateau as the equilibrium modulus. Through, $G_e(\varepsilon) = G_0 \varepsilon^z$ long-time elastic plateau associated post-gel can be seen to be attaining higher values with increase in $\varepsilon$ in the post-gel state.

An interesting outcome of the fitting procedure is that the state Winter and coworkers originally identified as the critical gel was well fit by the pre-gel expression (Eq. (21)) with $\varepsilon = 0.003$ rather than by the power law associated with the critical gel ($\varepsilon = 0$). Interestingly, the associated curve is nearly indistinguishable from a power law over most of the time range, which may explain why it was originally interpreted as the critical gel state. Nonetheless, the pre-gel expression with $\varepsilon = 0.003$ captures a subtle but systematic downward curvature at the longest accessible times, which the pure power law cannot reproduce, resulting in a superior fit. This example also illustrates a general challenge in the experimental identification of the critical gel state. Since $\tau_{max,S}$ diverges as $\varepsilon^{-1/\kappa}$, the states that are very close to but not at the gel point may appear indistinguishable from the critical gel over any finite experimental window. The present framework, which allows explicit parameterizing of the distance from the critical point, therefore, provides the sensitivity needed to make this distinction.

The fit shown in Fig. 1(a) has four modes ($N = 4$) to ensure the best fit. In Fig. 1(b) we plot the mechanistic decomposition of the four-mode fit for the pre-gel state at $\varepsilon = 0.0653$. The figure shows the individual contributions of each of the four terms in the summation of Eq. (21) alongside their sum and the experimental data. Each individual mode has a form $w_{S,l} S t^{-n} \exp(-B_{S,l}\varepsilon t^{\kappa})$, but with different values of $w_{S,l}$ and $B_{S,l}$. At short times, all modes share the same power-law slope of $n$. As time progresses, the curves deviate from this common envelope, reflecting the corresponding successively slower relaxation processes (decreasing rate coefficients $B_{S,l}$). The summation of all four modes yields a continuous curving relaxation modulus curve that matches the experimental data very well. The hierarchy of fits shown in Fig. 1(b) and all the fits shown in Fig. 1(a) that employ identical fitting parameters therefore emphasize that real crosslinking networks possess a distribution of multiple relaxation environments that a single mode cannot capture.

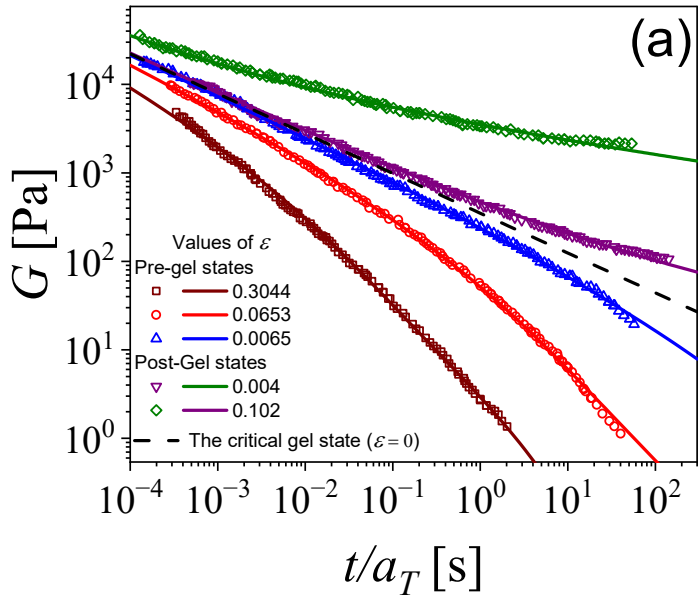

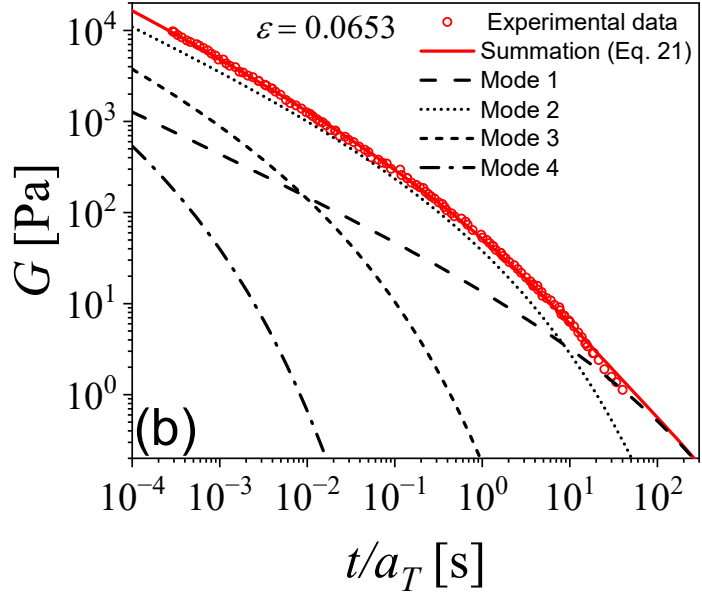

**Fig. 1.** A fit of Eqs. (21) and (46) to the relaxation modulus data across the sol-gel transition for a crosslinking PDMS system. (a) Stress relaxation modulus $G(t/a_T)$ is plotted as a function of reduced time $t/a_T$ for three pre-gel and two post gel states. Symbols represent experimental data obtained from the crosslinking PDMS system of Winter and coworkers. Solid lines are simultaneous fits of the pre-gel expression Eq. (21) and the post-gel expression Eq. (46) to all datasets using the unified parameter set listed in Table 1 ($N = 4$ modes). The critical gel state, shown as a dashed line, follows a pure power law $G = St^{-n}$. (b) Decomposition of the four-mode pre-gel fit (Eq. (21) ) into individual mode contributions for $\varepsilon = 0.0653$. Open circles represent experimental data, while the solid red line represents the sum over all four modes. The corresponding contributions of the four modes, each in the form $w_{S,l} St^{-n} \exp(-B_{S,l}\varepsilon t^{\kappa})$, are also plotted as lines.

**Table 2.** Model parameters for the pre-gel and post-gel relaxation modulus expressions used in Fig.1.

| Parameter | Value |
| --- | --- |
| Gel strength ($S$) | 350.166 Pa s$^n$ |
| Power-law exponent ($n$) | 0.45 |
| Gel-kinetics exponent ($\kappa$) | 0.29 |
| Equilibrium modulus prefactor ($G_0$) | 3677.22 Pa |
| Gel-point exponent ($z = n/\kappa$) | 1.55 |
| Number of relaxation modes ($N = M$) | 4 |

| Modes ($l$) | $B_{S,l} = B_{G,l}\ [s^{-\kappa}]$ | $w_{S,l} = w_{G,l}$ |
|---|---|---|
| 1 | 6.55 | 0.0592 |
| 2 | 25.03 | 0.5520 |
| 3 | 95.67 | 0.2619 |
| 4 | 365.60 | 0.1269 |

We now analyse the experimental dynamic moduli data with respect to the expressions derived for the pre-gel and post-gel states. For this purpose, we use the experimental data on an aqueous poly(vinyl alcohol) (PVA) solution studied by Joshi et al.[27] The aqueous PVA solution system differs from the chemically crosslinking systems that have dominated the sol-gel transition literature: it forms thermoreversible physical crosslinks. In this system, the junction zones between PVA chains are composed of microcrystalline domains aided by hydrogen bonding, which form upon cooling and dissolve upon heating.[4, 27, 31] Consequently, the system traverses the critical gel state reversibly in either direction. The data in Fig. 2 correspond to the gel-to-sol transition, obtained by cooling the PVA solution system through the critical temperature $T_c$=20°C. In Fig. 2, we plot $G'$ and $G''$ as function of temperature $T$ at five representative frequencies spanning nearly three decades. On the pre-gel side ($T > T_c$) as well as the post gel side ($T < T_c$) the theoretical expressions derived in Section II (For pre-gel Eqs. (28) and (29)) and for post-gel Eqs. (50) and (51)) describe the experimental data very well across all frequencies for a single mode expression. The fitted parameters are mentioned in Table 3, out of which only the values of $B_{S,l}\ (= B_{G,l})$, $G_0$ and $\varepsilon$ were fitted to the experimental data, all the other values were taken from the analysis of the experimental data performed by Joshi et al.[27] As mentioned before, only the product $\varepsilon B_{S,l}$ matters for the fitting purposes, but since theoretically $\varepsilon$ depends on temperature, for convenience, we show both the values separately. Reference value of $\varepsilon$ at 18 and 22°C is taken to be 0.622 and scaled at different temperatures by considering $\varepsilon \sim |T - T_c|$. The quality of the fit for both the moduli on the pre-gel side is uniformly good across all the frequencies, and is equally good for $G'$ in the post-gel side. However, for the post-gel $G''$ the model seems to overpredict the values at lowest explored frequencies for the temperatures farthest from the critical gel state. This asymmetry in fit quality between the pre-gel and post-gel sides could be rooted in an intrinsic difference in the mathematical structure of the two expressions. While the pre-gel expression involves an infinite convergent series in $\varepsilon$, the post-gel expression is truncated and hence its accuracy could reduce as $\varepsilon$ grows. Interestingly, the crossover of $G'$ and $G''$ is also captured very well by the model, further validating the theoretical framework at the transition itself. Overall, Fig. 2 demonstrates that the theoretical framework developed in this work provides a quantitative description of the evolution of dynamic moduli across the sol-gel transition.

**Table 3.** Model parameters for the pre-gel and post-gel dynamic moduli expressions used in Figs. 2 and 3.

| Parameter | Value |
|---|---|
| Gel strength ($S$) | 1.184 $Pas^n$ |
| Power-law exponent ($n$) | 0.77 |
| Gel-kinetics exponent ($\kappa$) | 0.266 |
| Equilibrium modulus prefactor ($G_0$) | 1.436 Pa |
| Gel-point exponent ($z = n/\kappa$) | 2.9 |
| Reference value of $\varepsilon$ at 18 and 22°C | 0.622 |
| Number of relaxation modes ($N = M$) | 1 |

| Modes ($l$) | $B_{S,l} = B_{G,l}\ [s^{-\kappa}]$ | $w_{S,l} = w_{G,l}$ |
|---|---|---|
| 1 | 1.697 | 1 |

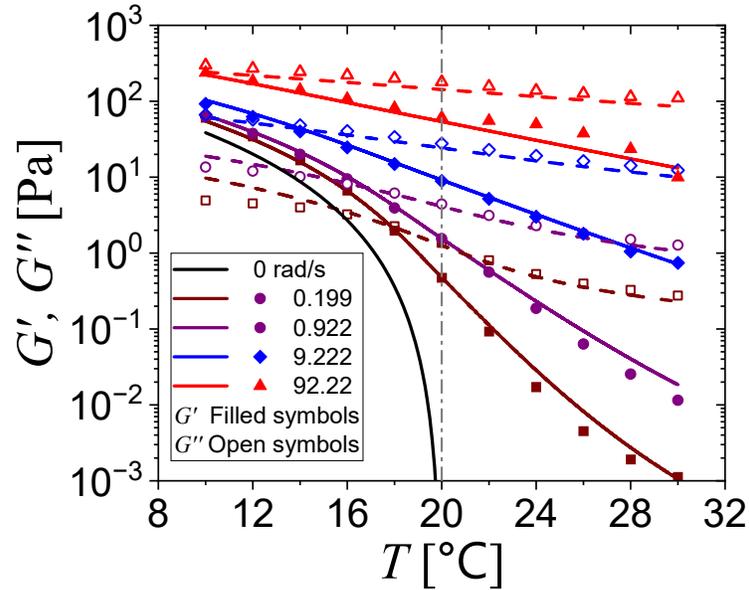

**Figure 2.** Storage modulus $G'$ (filled symbols) and loss modulus $G''$ (open symbols) are plotted as a function of temperature $T$ for an aqueous poly(vinyl alcohol) solution undergoing thermoreversible sol-gel transition, at five representative angular frequencies. The critical gel state is located at $T_c = 20$°C (vertical dash-dot line), with the pre-gel states corresponding to $T > T_c$ and the post-gel states occupying $T < T_c$. Experimental data is taken from Joshi et al.[27] Full lines ($G'$) and dashed lines ($G''$) are fits of the pre-gel expressions (Eqs. (28) and (29)) and the post-gel expressions (Eqs. (50) and (51)) using the single-mode ($N = M = 1$) parameter set listed in Table 3.

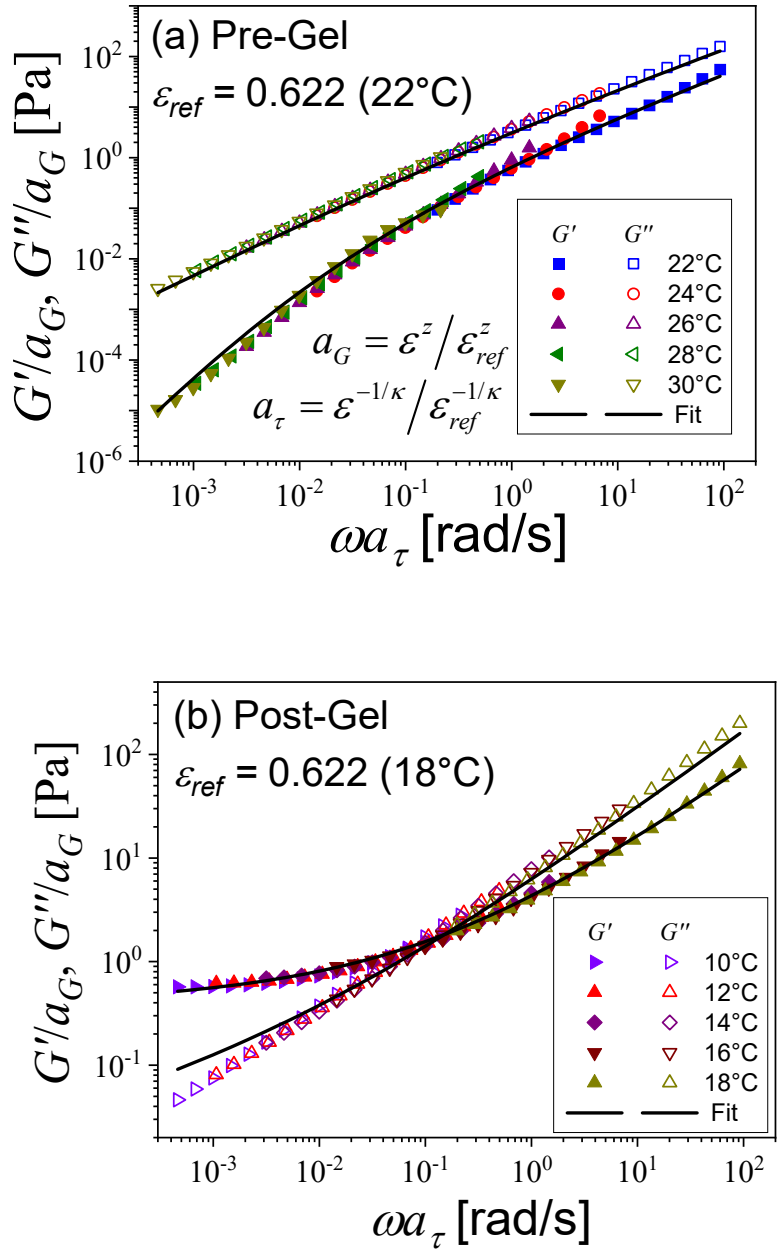

**Figure 3.** Master curves of the storage modulus $G'/a_G$ and loss modulus $G''/a_G$ are plotted as a function of the frequency $a_\tau \omega$ for an aqueous poly(vinyl alcohol) solution undergoing thermoreversible gel-to-sol transition upon cooling, for (a) the pre-gel state and (b) the post-gel state. The master curves are constructed from experimental dynamic moduli data at multiple temperatures reported by Joshi et al.,[27] using the frequency shift factor $a_\tau = \tau_{\max}(\varepsilon)/\tau_{\max}(\varepsilon_{\rm ref})$ and the modulus shift factor $a_G = G_e(\varepsilon)/G_e(\varepsilon_{\rm ref})$, with the scaling relations $\tau_{\max} \propto \varepsilon^{-1/\kappa}$ and $G_e \propto \varepsilon^z$ leading to the exact shift factors. The reference states are $\varepsilon_{\rm ref} = 0.622$ at $T = 22°C$ for the pre-gel and $\varepsilon_{\rm ref} = 0.622$ at $T = 18°C$ for the post-gel. The critical gel state corresponds to $T_c = 20°C$ (not shown; for details see Joshi et al.[27]). Solid lines are the single-mode ($N = M = 1$) fits from the pre-gel (Eqs. (28) and (29)) and post-gel (Eqs. (50) and (51)) expressions with parameters mentioned in Table 3. A possible origin of discrepancy between the data and the fit for post-gel $G''$ at low frequencies is discussed in the text.

In Fig. 3, we plot the pre-gel and post-gel master curves of frequency dependence of dynamic mdouli at the reference temperatures of $T$=22°C and $T$=18°C, respectively. It should be noted that the data shown in Fig. 3 are time-cure superposition master curves obtained from measurements at multiple temperatures. Joshi et al.[27] reported the scaling exponents $\kappa$ and $z$ for the PVA solution system but did not obtain the superposition. Here, those exponents are used directly to obtain the shift factors such that a frequency shift factor $a_\tau = \tau_{\max}(\varepsilon)/\tau_{\max}(\varepsilon_{\text{ref}})$ is used along the horizontal axis, while a modulus shift $a_G = G_e(\varepsilon)/G_e(\varepsilon_{\text{ref}})$ is employed along the vertical axis. Using $\tau_{\max} = \tau_S\, \varepsilon^{-1/\kappa}$ and $G_e = G_0 \varepsilon^z$, these shift factors take the explicit forms: $a_\varepsilon = \left(\frac{\varepsilon_{\text{ref}}}{\varepsilon}\right)^{1/\kappa}$, $a_G = \left(\frac{\varepsilon}{\varepsilon_{\text{ref}}}\right)^z$ leading to the plotted variables $a_\tau \omega = \omega\, \tau_{\max}(\varepsilon)/\tau_{\max}(\varepsilon_{\text{ref}})$ and $G'/a_G = G' G_e(\varepsilon_{\text{ref}})/G_e(\varepsilon)$. It can be seen that, with these shift factors, the dynamic moduli data at different temperatures collapse onto a single master curve, both in the pre-gel and post-gel states, providing a direct, quantitative validation of the scaling relations. In the pre-gel state, however, there does not exist a true equilibrium modulus. Despite that, we use $G_e(\varepsilon) = G_0 \varepsilon^z$ as a normalization scale because the hyperscaling relation $z = n/\kappa$ ensures that $G_0 \varepsilon^z \propto \varepsilon^{n/\kappa}$ has the same power-law dependence on $\varepsilon$ as the natural characteristic modulus: $S\tau_{max}^{-n} \sim \varepsilon^{n/\kappa}$. Since the two scales are proportional, either may serve as the vertical axis reference. The master curve construction, therefore, works self-consistently in both states precisely because the hyperscaling relation holds.

In Fig 3(a) and (b), the solid lines are the fit of Eqs. (28) and (29) for pre-gel and Eqs. (50) and (51) for the post-gel with single-mode ($N = M = 1$). In the pre-gel state, shown in Fig 3(a), the moduli scale associated with the superposition spans eight decades over the plotted frequency window, which is over five decades. At low frequencies $G'' \gg G'$, as expected from the pre-gel states that are viscoelastic liquids with no long-time or low frequency elasticity. At high frequencies, on the other hand, both moduli converge toward parallel power laws with slope $n = 0.77$, recovering the critical gel signature as the relaxation cutoff is left far behind on the frequency axis. The model, with a single mode, captures this entire behaviour. In the post-gel state (Fig (b)), the normalized $G'$ approaches the plateau $G_e(\varepsilon_{\text{ref}}) = 0.47$ Pa at low frequencies, while $G''$ tends to zero, suggesting an unambiguous elastic solid character of the percolated network. At high frequencies, both moduli again approach the power-law slope with $n = 0.77$. As is the case shown in Fig. 2, the model tends to overpredict the experimental behavior for $G''$ for the lowest frequencies. The corresponding data shown in the low frequency region primarily corresponds to temperature $T$ =10°C, which is the farthest explored temperature from the critical gel state with the highest $\varepsilon$. As discussed before, we believe that possibility of truncated nature of series for the post-gel regime limits its predictive capability with an increase in $\varepsilon$ especially for $G''$. This issue does not arise for $G'$ for the post-gel states, as the elastic contribution $G_0 \varepsilon^z$ provides an additional back-up, which is not available for $G''$. Nonetheless, considering the fit involves capturing the four sets of data simultaneously, each for over five decades of frequencies, with a single set of parameters, the theory describes the response very well.

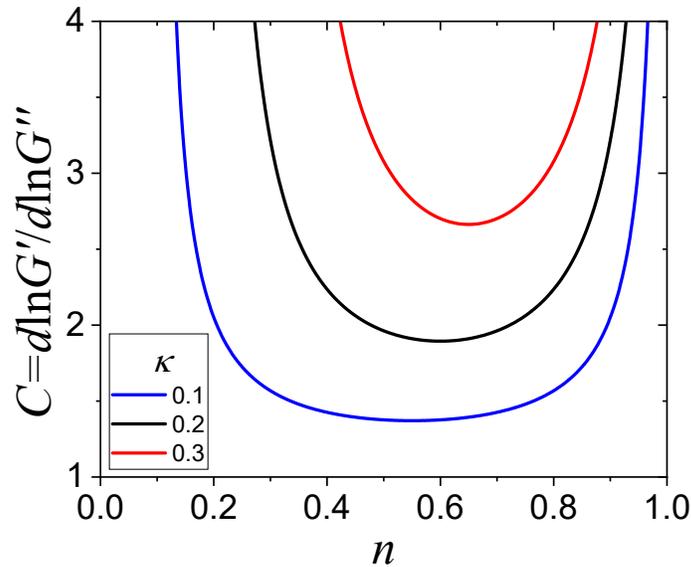

**Figure. 4.** The parameter $C$ ($= \partial lnG'/\partial lnG''$ as $\varepsilon \to 0$) given by Eqs. (58) is plotted as a function of the critical relaxation exponent $n$ for three values of the relaxation scaling exponent $\kappa$ = 0.1, 0.2, and 0.3. The region $n < \kappa$ is physically inaccessible as discussed in Section II.4 suggesting the derivative $\partial G'/\partial G''$ to have only positive values at the critical gel state.

In Fig. 4, we plot analytically derived parameter $C$ as a function of $n$ for three values of $\kappa$ that span the experimentally reported range. As shown by Eq. (58), the expression for $C$, depends only on $n$ and $\kappa$, and is independent of frequency, gel strength $S$, as well as the other parameters of the relaxation spectrum. The fact that $C$ is frequency and mode independent is a non-trivial result, which suggests that the value of $C$ an intrinsic fingerprint of a material. Across various gel forming systems, experimentally a value of $n$ is observed to be bounded by: $0.2 \lesssim n \lesssim 0.9$. Considering, $\kappa$ to lie between 0.1 and 0.3, Eq. (58) and Fig. 4 suggest that the parameter $C$ remains approximately in the range 1.5 to 4, which provides a natural explanation for why the mean value of $C$ has been reported to be close to 2 across a wide range of chemically and physically crosslinking systems. The present theoretical framework, for the first time, shows that this much-discussed empirical parameter $C$ is precisely computable and why its experimentally reported values lie between 1.5 and 4. Furthermore, Fig. 4 and Eq. (58) make it clear that $C$ is positive throughout the physically accessible parameter space. This is a direct consequence of the constraint $n > \kappa$ established in Section II.4. Importantly, the absence of experimental reports of $n \leq \kappa$ is not coincidental but is a necessary consequence of the underlying physics. Fig. 4 makes this constraint explicit by the inaccessibility of the lower-left region of the parameter space. This consequent constraint on $C$ leads to an important physical interpretation. Since $C = \partial lnG'/\partial lnG''$ as $\varepsilon \to 0$, a positive value of the same means that as a material passes through the critical gel state, $G'$ and $G''$ must change in the same direction with respect to $\varepsilon$. In other words, if $G'$ increases (decreases) as the material approaches the critical gel state from the pre-gel (post-gel) side, $G''$ must also increase (decrease). This co-directionality of $G'$ and $G''$ evolution must hold for any system satisfying $n > \kappa$, and it is consistent with all experimental observations on the sol-gel transition reported to date. A negative value of $C$, which would imply that $G'$ and $G''$ evolve in opposite directions as the critical state is approached, is therefore, not merely unusual but physically forbidden within the framework of continuous sol-gel transitions governed by percolation-type critical phenomena.

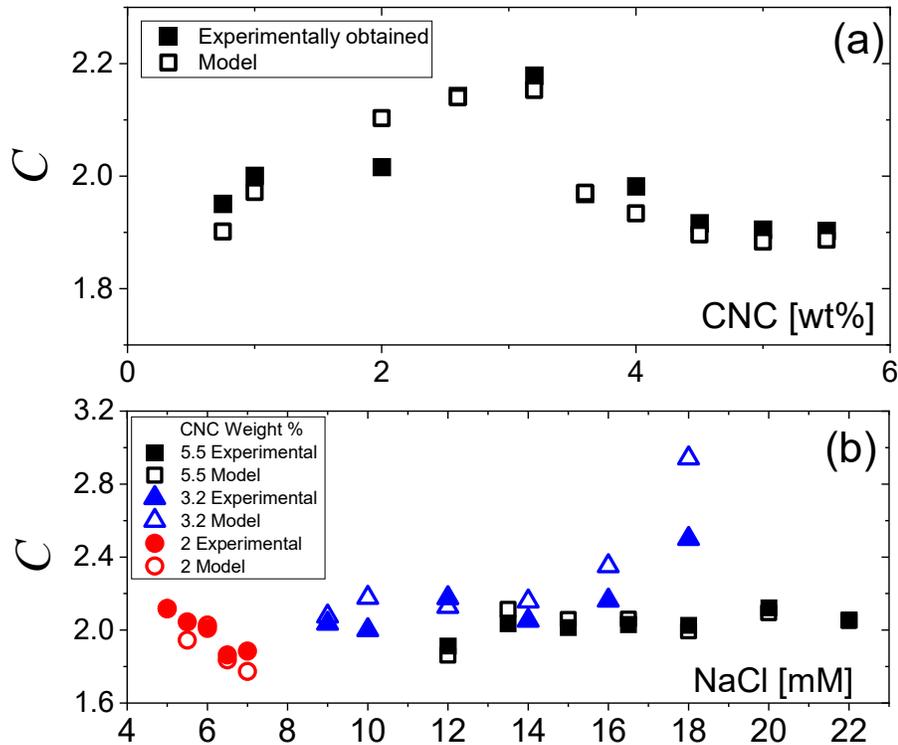

**Figure 5.** Comparison of experimentally measured values of the parameter $C$ for cellulose nanocrystal (CNC) suspensions from Morlet-Decarnin et al.[30] with those theoretically predicted by the model. The parameter $C$ is plotted as a function of (a) CNC concentration (at varying salt concentrations and (b) NaCl concentration for three CNC weight fractions. Filled symbols represent experimental values of $C$ reported by Morlet-Decarnin et al.,[30] while open symbols denote theoretical predictions computed from Eq. (58), using the experimentally extracted values of $n$ (equivalent to $\beta$ in Morlet-Decarnin et al.[30]) and $\kappa$ (equivalent to $\kappa'$ in Morlet-Decarnin et al.[30]) reported in the same work.

We now compare the predictions of Eq. (58) directly with the experimental data. Fig. 5 provides a rigorous test of the analytical expression for the parameter $C$ (Eq. (58)), wherein we plot an independent experimental dataset from the recent work of Morlet-Decarnin et al.[30] on charged cellulose nanocrystal (CNC) suspensions. In this work, CNC gels were prepared over a wide range of CNC concentrations (2 – 5.5 wt%) and NaCl concentrations (5 – 22 mM). Morlet-Decarnin et al.[30] independently estimate the critical relaxation exponent (denoted $\beta$ in the original paper, equivalent to $n$ in the present work), the dynamic critical exponent (denoted $\kappa'$ in the original paper, equivalent to $\kappa$ on the present work) and parameter $C$ from time-resolved linear viscoelastic experiments. In Fig. 5 we plot parameter $C$, which they obtain experimentally and the parameter $C$ predicted by the model by substituting experimentally obtained parameters $n$ and $\kappa$ by Morlet-Decarnin et al.[30] in Eq. (58), without any additional fitting. It can be seen that the agreement between the theoretical predictions and the experimental data is remarkably good across all explored concentrations of CNC as well as the salt, as shown in Figs. 5(a) and 5(b). This figure also confirms an important result of Section II, that the parameter $C$, which also denotes $\left(\frac{\partial \ln G'(\omega,\varepsilon)}{\partial \ln G''(\omega,\varepsilon)}\right)_{\varepsilon \to 0}$, is an intrinsic material parameter that depends solely on $n$ and $\kappa$ and is always necessarily positive.

The theoretical framework developed in this paper opens several directions for future investigation. To begin with, there is a need to identify gel-forming materials that possess very small critical exponents $n < 0.2$. Such systems are expected to be highly elastic gels with significantly large crosslink density (and hence fractal dimension) at the critical gel state. This regime of $n < 0.2$ is not much explored in the literature. After identifying such a system, it is important to estimate $\kappa$ for such a system that would provide a direct test of the constraint $n > \kappa$ established in Section II.4. If materials with $n \leq \kappa$ are indeed found to undergo continuous sol-gel transitions, they would represent genuinely new physics beyond the present framework. Conversely, if no systems show $n \leq \kappa$ as proposed in this work, and values of $n$ lower than 0.2 indeed exist, it would suggest that $\kappa$ could be a function of $n$, in the low $n$ limit, such that as $n$ decreases $\kappa = \kappa(n)$ adjusts to maintain $n > \kappa$ across all physical systems. We believe that such a possibility itself warrants systematic theoretical and experimental investigation. Another closely related question is, what is the practically possible range of values that $\kappa$ can take. We know that the critical relaxation exponent $n$ varies over a much broader region, and along with $S$ completely characterizes the critical gel state. However, the relaxation scaling exponent $\kappa$ is an equally important parameter that determines how the critical state is approached as gelation transition occurs in either direction. It is, therefore, important to investigate the bounds on the same. Furthermore, the present work, as well as most of the literature on the gelation transition, treats $n$ as a constant throughout the transition. This is also consistent with ideal percolation behavior and has been validated for both crosslinking polymers as well as colloidal systems. However, Mours and Winter[21] reported a unique system having high molecular weight entangled precursors, wherein $n$ does change with extent of crosslinking. Extending the present framework to accommodate systems with varying $n$ varies would require generalizing the relaxation modulus to $G(t,\varepsilon) = S(\varepsilon)t^{-n(\varepsilon)}\Phi(t/\tau_{max,S})$. Moreover, meaningful continuity conditions must be formulated at the critical point. Such an extension would connect the current percolation-based framework to the mean-field regime predicted by de Gennes for high-molecular-weight vulcanizing systems.[72] Overall, the results presented in this section demonstrate that the theoretical framework developed in Section II provides a sound description of the experimental data on the evolution of linear viscoelastic properties across the sol-gel transition. We show that a single set of material parameters and mode coefficients simultaneously describe the relaxation modulus as well as the dynamic moduli. However, as mentioned, the specific forms of the relaxation modulus and dynamic moduli proposed in this work are not unique. There may exist other forms (not necessarily analytical) that while obeying the intrinsic constraints may also lead to various linear viscoelastic parameters associated with the gelation transition. We believe that various open issues discussed herein will further strengthen our understanding of the approach to the critical state during the gelation transition.

**IV. Conclusions**

This work presents a comprehensive theoretical framework for the evolution of linear viscoelastic properties across the sol-gel transition, which leads to several significant and previously unestablished results. To begin with, we derive a general expressions for the relaxation modulus $G(t,\varepsilon)$ and the corresponding dynamic moduli $G'(\omega,\varepsilon)$ and $G''(\omega,\varepsilon)$ in both the pre-gel and post-gel states. We do not assume any specific functional form beyond the physical constraints of causality, monotonic stress relaxation, and the experimentally observed limiting behavior as the critical gel state is approached. We propose that the pre-gel relaxation modulus has a multi-mode exponential series form. The analogous post-gel expression, on the other hand, is a truncated polynomial series whose termination is necessitated by the requirement of monotonic relaxation. Both these forms converge to the Winter–Chambon power-law expression as the distance from the critical gel state vanishes. We validate these expressions simultaneously against relaxation modulus data for a chemically crosslinked PDMS system undergoing a sol-gel transition and dynamic moduli data for a thermoreversible polyvinyl alcohol gel system undergoing a gel-sol transition.

A central and conceptually important result of this work is the demonstration of the symmetry condition $\kappa_S = \kappa_G$, which has previously been treated as an empirical experimental observation in the literature. We show that this condition is a necessary consequence of the continuity of derivatives of the dynamic moduli with respect to the degree of crosslinking at the critical gel point. Remarkably, since $\kappa_S = (1-n)/s$ and $\kappa_G = n/z$, this equality in turn implies the validity of the hyper-scaling relation $n = z/(z+s)$, which correlate the critical relaxation exponent ($n$) to the viscosity and equilibrium modulus exponents (respectively $s$ and $z$). We, thus, show that the validity of this hyperscaling relation is not an independent empirical observation but a direct consequence of physical continuity at the gel point. Very importantly, this work therefore proposes that experimental verification of the hyperscaling relation across diverse material systems is required to provide a consistency check for the continuity of the transition.

The next important result, proposed by the present theoretical frameworks for the first time, is that the regime $n < \kappa$ is physically inaccessible for any system undergoing a continuous gelation transition. We show that for $n < \kappa$, the truncated post-gel series collapses to its zeroth-order term, rendering the relaxable component of the relaxation modulus independent of $\varepsilon$ and making it impossible to satisfy the continuity condition at the critical gel point. This requirement is also reinforced by a proposal due to de Gennes restricting the equilibrium modulus exponent $z$ above unity for a three-dimensional percolating network.[68] The constraint $n > \kappa$, therefore, simultaneously constitutes a lower bound on $n$ for systems with a given $\kappa$. We also obtain the parameter $C = \left(\frac{\partial \ln G'(\omega,\varepsilon)}{\partial \ln G''(\omega,\varepsilon)}\right)_{\varepsilon \to 0}$, analytically for the first time as $C = \cot\left(\frac{(n-\kappa)\pi}{2}\right)\tan\left(\frac{n\pi}{2}\right)$, which suggests that $C$ depends only on $n$ and $\kappa$, and is independent of frequency, gel strength, and all mode-specific parameters. This result also explains the long-standing observation in the literature that $C$ takes values close to 2 (between 1.5 and 4) across a wide range of chemically and physically crosslinking (gel-forming) systems. Within the experimentally reported range of $n$ and $\kappa$: $0.2 \lesssim n \lesssim 0.9$ and $0.1 \leq \kappa \leq 0.3$, our theoretical framework elegantly predicts near-constant value of $C$. The analytical expression further establishes that $C$ is strictly positive throughout the physically accessible parameter space $n > \kappa$, which implies that $G'$ and $G''$ must always evolve in the same direction as the system passes through the critical gel state.

Overall, the proposed theoretical framework and its validation of the experimental data establish that the linear viscoelastic evolution across the gelation transition is governed by a tightly constrained and internally consistent theoretical structure. It is like a zigzag puzzle that leaves a little room for variation, leading to symmetry relaxation dynamics across the transition, hyper-scaling relations, the inaccessibility of region $n<\kappa,\kappa$ and the near-universal value of $C$. All these results emerge as necessary consequences of a single unifying physical requirement: the continuity of the linear viscoelastic properties and their derivatives as the material transitions through the critical gel point.

## Appendix A: Derivation of the Scaling and Hyperscaling Relations Near the Sol-Gel Transition

The sol-gel transition is a critical phenomenon in which a liquid-like sol transforms into a solid-like gel. The specific transition point is characterized by the degree of crosslinking expressed in terms of reduced distance from critical state $\varepsilon = |p - p_c|$. As the transition is approached, various material properties exhibit power-law scaling behaviors. This appendix provides a derivation, based on scaling arguments, of the fundamental scaling and hyperscaling relations that govern the viscoelastic behavior on both sides of the critical gel point.

### A.1 Pre-gel State $(p_c - p = \varepsilon < 0)$

In the sol state, two key properties diverge as the critical gel point is approached. The viscosity scales as $\eta_0 \sim \varepsilon^{-s}$, where $s$ is the viscosity divergence exponent. The corresponding longest relaxation time $(\tau_{max,S})$ scales as: $\tau_{max,S} = \tau_S \varepsilon^{-1/\kappa_S}$, where $\kappa_S$ is the pre-gel relaxation time exponent and $\tau_S$ is a material-dependent prefactor. We also know that at the critical gel point ($\varepsilon = 0$), for any value of $n$ over the entire range ($0 < n < 1$) and for times $t_0 < t < \infty$ we have: $G(t) = St^{-n}$, where $S$ is the gel strength. In the pre-gel state, there is a power-law is cut off at the longest relaxation time $\tau_{max,S}$ such that:

$$G(t) = St^{-n} \text{ for } t_0 < t \ll \tau_{max,S}$$

$$G(t) = 0 \text{ for } t \gg \tau_{max,S}$$
A1

Knowledge of relaxation modulus leads to the viscosity given by:

$$\eta_0 = \int_0^\infty G(t)\, dt \approx \int_{t_0}^{\tau_{max,S}} S\, t^{-n} dt = \frac{S}{1-n}[\tau_{max,S}^{1-n} - t_0^{1-n}] \sim \tau_{max,S}^{1-n}. \qquad A2$$

While deriving Eq. (A2), we assume that relaxation modes faster than $t_0$ ($\ll \tau_{max,S}$) contribute negligibly to the viscosity. Substituting Eq. (5) into Eq. (A2) gives

$$\eta_0 \sim \tau_{max,S}^{1-n} \sim \varepsilon^{-(1-n)/\kappa_S}. \qquad A3$$

Comparing Eq. (4) with Eq. (A3), therefore leads to:

$$\kappa_S = \frac{1-n}{s}. \qquad \text{A4}$$

For the general expression of relaxation modulus in the pre-gel state (Eq. (21)) derived in this work,

$$\eta_0 \approx \int_{t_0}^{\infty} S t^{-n} \sum_{l=1}^{N} w_{S,l} \exp(-B_{S,l}\varepsilon t^{\kappa_S}) dt = \frac{S}{\kappa_S} \sum_{l=1}^{N} w_{S,l} (B_{S,l}\varepsilon)^{\frac{n-1}{\kappa_S}} \Gamma\left(\frac{1-n}{\kappa_S}, B_{S,l}\varepsilon t_0^{\kappa_S}\right), \qquad \text{A5}$$

where $\Gamma(z, x)$ is the upper incomplete Gamma function, given by $\Gamma(z,x) = \int_{x}^{\infty} t^{z-1} e^{-t} dt$. For a limiting case of $t_0$ being sufficiently small (that means the contribution of modes faster than $t_0$ is negligible), the integral given by Eq. (A2) converges leading to:

$$\eta_0 \approx \frac{S}{\kappa_S} \Gamma\left(\frac{1-n}{\kappa_S}\right) \left[\sum_{l=1}^{N} w_{S,l} (B_{S,l})^{\frac{n-1}{\kappa_S}}\right] \varepsilon^{\frac{n-1}{\kappa_S}}, \qquad \text{A6}$$

which rigorously leads to the pre-gel scaling relation Eq. (A4). Another important integration is: $\int_{0}^{\infty} t\, G(t,\varepsilon)\, dt$, which can be estimated easily with the knowledge of relaxation modulus and is given by:

$$\int_{0}^{\infty} t\, G(t,\varepsilon)\, dt = \frac{S}{\kappa_S} \Gamma\left(\frac{2-n}{\kappa_S}\right) \sum_{l=1}^{N} w_l (B_{S,l}\varepsilon)^{-\frac{2-n}{\kappa_S}}. \qquad \text{A7}$$

Knowledge of this integration leads to the steady state compliance ($J_e^0$), which is a measure of the total recoverable (elastic) strain ($\gamma_\infty$) in a viscoelastic material in a limit of steady viscous state under a constant stress ($\sigma_0$), resulting in $J_e^0 = \gamma_\infty/\sigma_0$.[73] Increase in $J_e^0$, therefore, indicates that the material behaves more elastically before reaching steady flow. The expressions of $J_e^0$ in the sol state is given by:[73]

$$J_e^0 = \frac{\int_0^\infty t\, G(t,\varepsilon)\, dt}{\left[\int_0^\infty G(t,\varepsilon)\, dt\right]^2} = \left[\frac{\kappa_S}{S} \frac{\Gamma\left(\frac{2-n}{\kappa_S}\right) \sum_{l=1}^{N} w_l (B_{S,l})^{-\frac{2-n}{\kappa_S}}}{\left[\Gamma\left(\frac{1-n}{\kappa_S}\right) \sum_{l=1}^{N} w_l (B_{S,l})^{-\frac{1-n}{\kappa_S}}\right]^2}\right] \varepsilon^{-n/\kappa_S} \sim \varepsilon^{-z} \qquad \text{A8}$$

Considering that the hyper-scaling relation is validated, the last term suggests that, according to the present work, the steady state compliance scales as: $J_e^0 \sim \varepsilon^{-z}$, suggesting recoverable strain accumulated before reaching the steady state increases with $\varepsilon$.

**A.2 Post-gel State** $(p - p_c = \varepsilon < 0)$

Beyond the critical gel point, a space-spanning network percolates through the system, and an equilibrium (elastic) modulus emerges: $G_e(\varepsilon) = G_0 \varepsilon^z$, where $z$ is the elastic modulus exponent and $G_0$ is a material-specific constant. However, the finite clusters embedded within the network are free to relax, with a characteristic maximum relaxation time: $\tau_{\max,G} = \tau_G \varepsilon^{-1/\kappa_G}$, where $\kappa_G$ is the post-gel relaxation time exponent and $\tau_G$ is a constant. Typically, after an application of step strain, the structural length scales, from monomers to the infinite network, respond, resulting in maximum stress in the system. At early times ($t_0 < t < \tau_{\max,G}$), the system is still within the critical region, as it has not yet sensed the percolated network. Owing to their self-similar distribution of cluster sizes at the critical point, the relaxation of clusters occurs as per their characteristic length-scales, causing the stress to decay as a power law given by:

$$G(t) = St^{-n} \text{ for } t_0 < t \ll \tau_{\max,G}. \tag{A9}$$

At $t \approx \tau_{\max,G}$, the largest finite clusters undergo complete relaxation, which we can represent as the terminal time scale for the "sol" fraction in the post-gel state. At sufficiently large times ($t \gg \tau_{\max,G}$), all finite clusters have relaxed, and only the infinite network sustains the stress, resulting in a constant modulus $G_e$ given by:

$$G(t,\varepsilon) \sim \varepsilon^z, \text{ for } t \gg \tau_{\max,G}. \tag{A10}$$

The Eqs. (A7) and (A8) suggest that at a characteristic time $t \approx \tau_{\max,G}$ the relaxing component (following a power-law) must smoothly transition to the elastic plateau. Consequently, the magnitude of the power-law relaxation at $t = \tau_{\max,G}$ should be comparable to the equilibrium modulus:

$$G_{\text{relax}}(\tau_{\max,G}) \sim G_e(\varepsilon). \tag{A11}$$

If we drop the constant prefactors in this scaling argument, for $G_{\text{relax}}(\tau_{\max,G}) \sim \tau_{\max,G}^{-n}$ and $G_e(\varepsilon) \sim \varepsilon^z$, Eq. (A9) leads to: $\tau_{\max,G}^{-n} \sim \varepsilon^z$. However, since $\tau_{\max,G} = \tau_G \varepsilon^{-1/\kappa_G}$, we get a relationship between the power law coefficients in the post-gel state as:

$$\kappa_G = \frac{n}{z}. \tag{A12}$$

However, as discussed in this work, constraints associated with the continuity of the dynamic moduli and their derivatives, enforce $\kappa_S = \kappa_G$ resulting in a hyperscaling relationship:

$$n = \frac{z}{z+s}, \tag{A13}$$

for the sol-gel transition.

**Acknowledgment:** The author acknowledges financial support from the Science and Engineering Research Board, Government of India (Grant numbers: CRG/2022/004868 and JCB/2022/000040). The author also thanks Mr. Rahul Yadav for his assistance with Python programming. The author also thanks Prof. Sachin Shanbhag for fruitful discussions.